\documentclass[12pt]{article}
\usepackage{a4wide,epsfig}
\usepackage{scalefnt}
\usepackage[latin1]{inputenc}
\usepackage{amsmath}
\usepackage{amsfonts}
\usepackage[english]{babel}

\def\lQ{\Lambda_{\rm QCD}}
\newcommand{\nn}{\nonumber}
\newcommand{\be}{\begin{equation}}
\newcommand{\ee}{\end{equation}}
\newcommand{\bea}{\begin{eqnarray}}
\newcommand{\eea}{\end{eqnarray}}

\def\als{\alpha_{\rm s}}
\def\siml{{\ \lower-1.2pt\vbox{\hbox{\rlap{$<$}\lower6pt\vbox{\hbox{$\sim$}}}}\ }} 

\newcommand{\lp}{\left(}
\newcommand{\rp}{\right)}

\newcommand{\MS}{\overline{\rm MS}}

\begin{document}

\vspace*{1cm}
\begin{center}
  {\sc \large  $1/N_c$ and $1/n$ preasymptotic corrections to Current-Current correlators} \\
   \vspace*{2cm} {\bf Jorge~Mondejar$^a$
and Antonio~Pineda$^b$}\\
\vspace{0.6cm}
{\it $^a$\ Dept. d'Estructura i Constituents de la Mat\`eria\\
U. Barcelona, Diagonal 647, E-08028 Barcelona, Spain
\\
[10pt]
$^b$\ Grup de F\'\i sica Te\`orica and IFAE, Universitat
Aut\`onoma de Barcelona, E-08193 Bellaterra, Barcelona, Spain\\}
  \vspace*{2.4cm}
  {\bf Abstract} \\
 \end{center}
We obtain the 
corrections in $1/n$ and in $1/\ln n$ 
($n$ is the principal quantum number of the bound state) 
of the decay constants of scalar and pseudoscalar currents in two 
and four dimensions in the large $N_c$. We obtain them from the 
operator product expansion provided a model for the large $n$ mass spectrum is given. 
In the two-dimensional case the spectrum is known and the corrections obtained in this 
paper are model independent. We confirm these results by confronting them with the numerical 
solution of the 't Hooft model. We also consider a model at finite $N_c$ and obtain the 
associated decay constants that are consistent with perturbation theory.
This example shows that that 
the inclusion of perturbative corrections, or finite $N_c$ effects, to the OPE 
does not constrain the slope of the Regge trajectories, which remain a free parameter 
for each different channel.
\\[2mm]

\vspace*{5mm}
\noindent

\newpage


\section{Introduction}

Since the pioneering work of Ref. \cite{OPE}, 
the operator product expansion (OPE) has been intensively used in order 
to improve our understanding of the non-perturbative dynamics of the hadronic spectrum 
and decays for large excitations \cite{Blok:1997hs, Shifman:2000jv, Beane:2001uj, Golterman:2001pj, Cohen:2002st, Golterman:2002mi, 
Afonin:2004yb, Sanz-Cillero:2005ef, Shifman:2005zn, Cata:2006fu, Afonin:2006da, Golterman:2006gv, 
Jamin:2008rm}. Specially intense has been the study of the 
vacuum polarization correlator with different local currents. 
Typically, the large $N_c$ limit \cite{hooft1} is considered in these 
analysis, since it allows one to have infinitely narrow resonances at 
arbitrarily large energies. 

Recently, this problem has been revisited in Ref. \cite{Mondejar:2007dz}. 
The aim of that paper was to go beyond previous analysis by means of the systematic 
incorporation of the perturbative corrections to the lowest 
order (parton) result, as well as the subleading terms in $1/Q^2$, in the OPE expression 
of the current-current correlator. In order to reproduce the OPE expression from the 
hadronic version of the current-current correlator, it was necessary to include corrections in $1/n$ ($n$ is the 
principal quantum number of the bound state) in the expressions of the mass spectrum and 
decay constants. 
By doing so, it was shown that (in the large $N_c$ limit) the 
combination of the OPE with the knowledge of the large $n$ mass spectrum  allows one to perform 
a systematic determination of the preasymptotic corrections in $1/n$ of 
the decay constants. Even more so, power-like $1/n$ corrections can only be incorporated in 
the masses and decay constants after the inclusion of the perturbative corrections in 
$\als$ in the OPE expression. 

One of the aims of this paper is to continue this line of research. 
We first consider the scalar/pseudo-scalar correlators (in Ref. \cite{Mondejar:2007dz} only the 
vector and axial-vector correlators were considered). Besides performing the 
phenomenological analysis, we will have to deal with non-trivial anomalous dimensions, 
which make the analysis significantly different. We then consider the 
't Hooft model \cite{hooft2} (the large $N_c$ limit of two-dimensional QCD). 
We compute for the first time the $1/n$ corrections to the decay constants 
in this model. We  find that our results agree quite well with those 
obtained from a numerical evaluation using the 't Hooft equation. 
This provides us with a non-trivial check of our analytical result.  
Finally, we also consider a model at finite $N_c$ and obtain the 
associated decay constants that are consistent with perturbation theory.
This model is inspired in the large $N_c$ limit, with small widths for low $n$ 
resonances, but still with the right analytic properties in the complex plane. 
The construction of such a model is not completely trivial. To our knowledge 
it has first been used in Ref. \cite{Blok:1997hs}. Here we find that this model survives 
the test of the inclusion of the perturbative logarithms of $Q^2$, though the decay 
constants are no longer trivially related with the perturbative expression of the imaginary part of the 
correlator. We also show that 
the inclusion of perturbative corrections, or finite $N_c$ effects, to the OPE 
does not constrain the slope of the Regge trajectories, which remain a free parameter 
for each different channel.

As we have mentioned above, in this paper we obtain the 
$1/n$ corrections to the decay constants in the 't Hooft model. 
This computation is actually one of the main results of this paper. It 
is model independent, since the mass spectrum is known. 
These corrections are an unavoidable ingredient to undertake the 
analysis of the preasympotic corrections to deep inelastic scattering or 
B decays in two dimensions. Those processes can be in principle be studied 
through the operator product expansion. In order to search for duality violations 
it is compulsory to know the values of the decay constants for large $n$.
We expect to profit from our results in the near 
future to improve over the analysis of Ref. \cite{Mondejar:2006ct}, 
as well as to perform a similar analysis for deep inelastic scattering \cite{prep}. 

\section{Scalar/pseudoscalar correlator in D=4 dimensions}
\label{sec4D}

The scalar/pseudoscalar correlator reads
\be
\label{correlatorSP}
\Pi_{S,P}(q;\mu)
\equiv
i\int d^Dx e^{iqx}\langle vac|T\left\{J_{S,P}(x)J^{\dagger}_{S,P}(0)\right\}|vac\rangle
\,,
\ee
where $J_S=\bar \psi_f\psi_f$ and $J_P=i\bar \psi_f\gamma_5\psi_f$. The dependence on the 
renormalization scale $\mu$ appears because the currents have anomalous dimensions. 
In order to shorten the notation, we will
use $X=S,P$ in what follows when the distinction is not important. Using dispersion 
relations the correlator can be written like
\be
\Pi_{X}(Q^2;\mu)=\int_0^{\infty}dt\frac{1}{t+Q^2}\frac{1}{\pi}\text{Im}\Pi_{X}(t;\mu) \ ,
\ee
where $Q^2=-q^2$ is the Euclidean momentum. In four dimensions the parton-model result gives
\be
\Pi_{X}(q;\mu) \sim Q^2\ln{\frac{Q^2}{\mu^2}} \ .
\ee
In order to avoid these divergences, one typically considers derivatives of the 
correlators like
\be
{\cal A}_{X}(Q^2;\mu)\equiv Q^2{d \over dQ^2}\frac{\Pi_X(Q^2;\mu)-\Pi_X(0;\mu)}{Q^2}=
Q^2\int_0^{\infty}dt\frac{1}{(t+Q^2)^2}\frac{1}{\pi t}\text{Im}\Pi_{X}(t;\mu)
\,,
\ee
or 
\be
{\cal B}_{X}(Q^2;\mu)\equiv\frac{Q^4}{2}\frac{d^2\Pi_{X}(Q^2;\mu)}{(d Q^2)^2}=Q^4\int_0^{\infty}dt\frac{1}{(t+Q^2)^3}\frac{1}{\pi}\text{Im}\Pi_{X}(t;\mu) \ .
\ee
Since we are working in the large $N_c$ limit, the spectrum  
consists of infinitely narrow resonances with mass $M_X(n)$,
 and these functions can be written 
in the following way
\be
{\cal A}_{X}(Q^2;\mu)=
Q^2\sum_{n=0}^{\infty}
\frac{F^2_{X}(n;\mu)}{M_X^2(n)}{1 \over (Q^2+M_{X}^2(n))^2} \ ,
\ee
where $F_X \equiv \langle vac|J_X(0)|n\rangle$, and
\be
\label{Bhadr}
{\cal B}_{X}(Q^2;\mu)=\frac{Q^4}{2}\frac{d^2\Pi_{X}(Q^2;\mu)}{(d Q^2)^2}
=
Q^4\sum_{n=0}^{\infty}
\frac{F^2_{X}(n;\mu)}{(M^2_{X}(n)+Q^2)^3} \ .
\ee
For definiteness, in this article we will work with the function ${\cal B}_{X}(Q^2;\mu)$.
For large positive $Q^2$, 
one may try to approximate this function by 
its OPE, 
which in the chiral and $N_c \rightarrow \infty$ limit has the following structure 
\bea
\nn
&&{\cal B}_{X, OPE}(Q^2;\mu)=
\left(1-\frac{3\ }{4}N_c\frac{\als(Q^2)}{\pi}
\ln\left(\frac{Q^2}{\mu^2}\right)
+\cdots
\right)
\left[\frac{N_cQ^2}
{16\pi^2}\left(1+{\cal O}(\als)
\right)\right.
\\
&&
\left.
-\frac{3}{22N_cQ^2}
\left(1+{\cal O}(\als)\right)\beta(\als(\mu))\langle vac |G^2(\mu)| vac \rangle
+{\cal O}\left(\frac{1}{Q^4}\right)\right] \ .
\label{BOPE}
\eea
The term $\beta \langle G^2\rangle$ above is renormalization group invariant and 
$\beta=\mu^2d\alpha/(d\mu^2)$.

In order to obtain the perturbative piece of this expression one can use 
\be
\text{Im}\Pi^{pert.}_{X}(t;\mu)=\frac{t}{8\pi}N_c{\tilde R}(t;\mu) \ ,
\ee
where the expression of ${\tilde R}(s)=1+{\cal O}(\als)$ can be found 
in Ref. \cite{Chetyrkin:1996sr,Baikov:2005rw} with four loop accuracy. The one-loop expression 
for the coefficient of the gluon condensate can be read from Ref. \cite{Surguladze:1990sp}.

It should be stressed that ${\tilde R}(s;\mu)$
is renormalization-scale dependent and it fulfills the following equation 
(it is understood that ${\tilde R}(s;\mu)$ is also computed in the 
$\MS$ scheme)
\be
\mu^2\frac{d}{d\mu^2}m_{\MS}^2(\mu){\tilde R}(s;\mu)=0 \ .
\ee
Therefore, $\text{Im}\Pi_{X}(t;\mu)$ does not have a physical meaning by itself 
(in physical processes it should appear associated to masses or equivalent). 

\subsection{Matching}
\label{matching}

High excitations of the QCD spectrum are believed to satisfy linear Regge 
trajectories:
$$
\displaystyle{\lim_{n \rightarrow \infty}} \frac{M_{X,n}^2}{n}= \; {\rm constant}
.$$ 

For generic current-current correlators, such behaviour 
is consistent with perturbation theory in the Euclidean region at leading order in 
$\als$ if the decay constants are taken to be ``constants", ie. independent 
of the principal quantum number $n$. 

In this paper, we would like to include power-like corrections in $\als$ and $1/Q^2$ 
in a systematic way. In order to do so we will 
have to consider corrections to the linear Regge trajectories as well as 
to the decay constants. We will follow the same procedure used in Ref. 
\cite{Mondejar:2007dz} for the vector and axial-vector channels. 
We will consider that the large $n$ expression for the mass spectrum can 
be organized within a $1/n$ expansion in a systematic way starting from the 
asymptotic linear Regge behaviour. In order to fix (and simplify) the problem we will assume 
that no $\ln n$ term appears in the mass spectrum\footnote{This is a simplification. If 
one considers, for instance, the 't Hooft model \cite{hooft2}, $\ln n$ terms do indeed appear, 
as we will see in the next section. If we relax this condition one can only fix the ratio between the decay constant and the 
derivative of the mass. Actually, this can be done in a model independent way. The explicit formulas are shown in the appendix.}.  
Therefore, we write the mass spectrum in the following way (for large $n$)
\be
\label{massn}
M_{X}^2(n)=\sum_{s=-1}^{\infty}B_X^{(-s)}n^{(-s)}=B_X^{(1)}n+B_X^{(0)}+\frac{B_X^{(-1)}}{n}+\cdots
\ee
where $B_X^{(-s)}$ are constants. We define $M_{X,LO}^2(n)=B_X^{(1)}n$, 
 $M_{X,NLO}^2(n)=B_X^{(1)}n+B_X^{(0)}$ and so on for the leading order (LO), next-to-leading order (NLO), etc. To shorten the notation, we 
will denote $B_X^{(1)}=B_X$, $B_X^{(0)}=A_X$ and $B_X^{(-1)}=C_X$.

For the decay constants, we will have a double expansion in $1/n$ and $1/\ln n$.  
\be
\label{decayn}
F_{X}^2(n;\mu)=n\sum_{s=0}^{\infty}F_{X,s}^{2}(n;\mu)\frac{1}{n^s}=n\lp F_{X,0}^{2}(n;\mu)+\frac{F_{X,1}^{2}(n;\mu)}{n}+\frac{F_{X,2}^{2}(n;\mu)}{n^2}+\cdots\rp
\,.
\ee
The logarithmic dependence on $n$ of the coefficients $F_{X,s}^{2}(n;\mu)$ has the following typical structure 
($\bar \gamma_0$ is defined in sec. \ref{secLO}):
\be
F_{X,s}^{2}(n;\mu)= \left(\frac{1}{\ln n}\right)^{2\bar \gamma_0}\sum_{r=0}^{\infty}C_{X,s}^{(r)}(\mu)\frac{1}{\ln^r n} \ .
\ee
As we did with the masses, we will define $F_{X,LO}^2(n;\mu)=F_{X,0}^{2}(n;\mu)$,  
$F_{X,NLO}^2(n;\mu)=F_{X,0}^{2}(n;\mu)+F_{X,1}^{2}(n;\mu)/n$, and so on. Note that 
in this case we also have an expansion in $1/\ln n$. 

We are now in the position to start the computation. Our aim is 
to compare the hadronic and OPE expressions of ${\cal B}_{X}(Q^2;\mu)$ 
within an expansion in $1/Q^2$, but keeping the logarithms of $Q$. 
In order to do so we have to arrange the hadronic expression appropiately. Our 
strategy is to split the sum over hadronic resonances into two pieces, for 
$n$ above or below some arbitrary but formally large $n^*$ such that 
$\lQ \ll \lQ n^* \ll Q$. The sum up to $n^*$ can be analytically expanded in $1/Q^2$ 
and will not generate $\ln Q^2$ terms (neither a constant term at leading 
order in $1/Q^2$). For the sum from $n^*$ up to $\infty$, we can use Eqs.  (\ref{massn}) and (\ref{decayn})
and the Euler-MacLaurin formula to transform the sum in an integral plus 
corrections in $1/Q^2$. Whereas the latter do not produce logarithms, the integral does. 
These logarithms of $Q$ are generated by the large $n$ behaviour of the bound 
states and the introduction of powers of $1/n$ is equivalent (once 
introduced in the integral representation, and for large $n$) to the 
introduction of (logarithmically modulated) $1/Q^2$ corrections in the OPE expression. 

Therefore, by using the Euler-MacLaurin formula, we write ${\cal B}_{X}(Q^2;\mu)$ in the
following way ($B_2=1/6$, $B_4=-1/30$, ...)
\bea
\label{BEuler}
\nn
{\cal B}_X(Q^2;\mu)&=&Q^4\int_0^{\infty} dn {F_X^2(n;\mu) \over (Q^2+M_X^2(n))^3}+
Q^4\left[\sum_{n=0}^{n^*-1}{F_X^2(n;\mu) \over (Q^2+M_X^2(n))^3}
- \int_0^{n^*}dn{F_X^2(n;\mu) \over (Q^2+M_X^2(n))^3}\right]
\\
&&
\left.
+
{Q^4 \over 2}{F_X^2(n^*;\mu) \over (Q^2+M_X^2(n^*))^3}
+
Q^4\sum_{k=1}^{\infty}(-1)^k{|B_{2k}| \over (2k)!}{d^{(2k-1)} \over dn^{(2k-1)}}
{F_X^2(n;\mu) \over (Q^2+M_X^2(n))^3}\right|_{n=n^*}
\,,
\eea
where $n^*$ stands for the subtraction point we mentioned above, such that 
for $n$ larger than $n^*$ one can use the asymptotic expressions (\ref{massn}) and 
(\ref{decayn}). This allows us to eliminate terms that vanish when 
$n \rightarrow \infty$. Note that the last sum in Eq. (\ref{BEuler}) is an asymptotic series, and 
in this sense the equality should be understood. 

Note also that for $n$ below $n^*$, we will not distinguish between LO, NLO, etc...
in masses or decay constants, since for those states we will not assume that one can do an expansion 
in $1/n$ and use Eqs. (\ref{massn}) and (\ref{decayn}).

Finally, note that the expressions we have for the masses and decay constants become more and 
more infrared singular as we go to higher and higher orders in the $1/n$ expansion. 
This is not a problem, since we always cut off the integral for $n$ smaller 
than $n^*$. Either way, the major problems in the correlator would come from the decay constants, since, in the case of 
the mass, $Q^2$ effectively acts as an infrared regulator.

And a final comment on renormalons: in principle, we are now in the position to match 
Eq. (\ref{BEuler}) with Eq. (\ref{BOPE}) order by order in $1/Q^2$. One should keep in mind however 
that each order in $1/Q^2$ of Eq. (\ref{BOPE}) suffers from renormalon ambiguities. Those ambiguities cancel between different orders
in the $1/Q^2$ expansion. A way to deal with this problem is to devise a scheme of subtracting renormalons from the 
perturbative series, passing the renormalon to the condensates, or to higher order terms in the $1/Q^2$ expansion, 
where the renormalon ambiguity cancels (see for instance \cite{Campanario:2005np} for an example of such a renormalon subtraction scheme). 
We will not do this explicitly in this paper, since it goes beyond our purposes and, 
with the precision we aim at here, these effects do not appear to be numerically 
dominant, at least for large $n$. Nevertheless, it remains to be seen whether they lead 
to some improvements for low $n$. 
 
\subsection{LO Matching}
\label{secLO}

We want to match the hadronic, Eq. (\ref{BEuler}), and OPE, Eq. (\ref{BOPE}), expressions for ${\cal B}_X(Q^2;\mu)$ at the 
lowest order in $1/Q^2$. Only the first term in Eq. (\ref{BEuler}) can generate logarithms or constant terms that are not suppressed by powers of $1/Q^2$, so this is the term that has to be matched to the perturbative part of the OPE expansion. We have to consider the lowest order expressions in $1/n$ for the 
masses and decay constants, i.e. $F_{X,LO}^2(n;\mu)$ and $M_{X,LO}^2(n)$,  
since the corrections in $1/n$ give contributions suppressed by powers of $1/Q^2$.
The matching condition is then
\be
{\cal B}_X^{pt.}\equiv Q^4\int_0^{\infty} dn {F_{X,LO}^2(n;\mu) \over (Q^2+M_{X,  LO}^2(n))^3}=\frac{N_cQ^2}{16\pi^2}\left(1+\frac{11-6\ \ln\left(\frac{Q^2}{\mu^2}\right)}{8}N_c\frac{\als(Q^2)}{\pi}+\dots\right) \ ,
\ee
which can be fulfilled by demanding
\be
\frac{n F^2_{X,LO}(n;\mu)}{|d M^2_{X,LO}(n)/dn|}=\frac{1}{\pi}\text{Im}\Pi_{X}^{pt.}(M^2_{X,LO}(n);\mu) \ .
\ee
This leads us to the following expression for $F^2_{(X),0}(n;\mu)$ 
with four-loop running precision 
\bea
\label{LO}
F^2_{X,0}(n;\mu)&=&\frac{B_{X}^2}{8\pi^2}N_c\left[\left(\frac{a(nB_{X}) }{a(\mu^2)}\right)^{\bar{\gamma_0}}\frac{c(a(nB_{X})}{c(a(\mu^2))}\right]^2\\
&&\times\left(1+r_1 a(nB_{X})+r_2 a(nB_{X})^2+r_3a(nB_{X})^3\right) \ ,\nn
\eea
where $a(\mu^2)=\als(\mu^2)/\pi$,
\bea
a(nB_{X})&=&a(\mu^2)\left\{1+a(\mu^2)\beta_0\ln\left(\frac{nB_{X}}{\mu^2}\right)\right.\\
&&+\frac{a(\mu^2)^3\left(\beta_0\ln\left(\frac{nB_{X}}{\mu^2}\right)+\bar{\beta_1}\ln\left(1+a(\mu^2)\beta_0\ln\left(\frac{nB_{X}}{\mu^2}\right)\right)\right)}{1+a(\mu^2)\beta_0\ln\left(\frac{nB_{X}}{\mu^2}\right)+a(\mu^2)\bar{\beta_1}\ln\left(1+a(\mu^2)\beta_0\ln\left(\frac{nB_{X}}{\mu^2}\right)\right)}(\bar{\beta_2}-\bar{\beta_1}^2)\nn \\
&&+a(\mu^2)\bar{\beta_1}\ln\left[1+a(\mu^2)\beta_0\ln\left(\frac{nB_{X}}{\mu^2}\right)+\frac{a(\mu^2)^3\beta_0\ln\left(\frac{nB_{X}}{\mu^2}\right)}{1+a(\mu^2)\beta_0\ln\left(\frac{nB_{X}}{\mu^2}\right)}(\bar{\beta_2}-\bar{\beta_1}^2)\right.\nn\\
&&\left.+a(\mu^2)\bar{\beta_1}\ln\left(1+a(\mu^2)\beta_0\ln\left(\frac{nB_{X}}{\mu^2}\right)+a(\mu^2)\bar{\beta_1}\ln\left(1+a(\mu^2)\beta_0\ln\left(\frac{nB_{X}}{\mu^2}\right)\right)\right)\right] \nn\\
&&\left.+\frac{2a(\mu^2)^4\beta_0\ln\left(\frac{nB_{X}}{\mu^2}\right)+a(\mu^2)^5\beta_0^2\ln^2\left(\frac{nB_{X}}{\mu^2}\right)}{\left(1+a(\mu^2)\beta_0\ln\left(\frac{nB_{X}}{\mu^2}\right)\right)^2}\left(\frac{\bar{\beta_1}^3}{2}-\bar{\beta_1}\bar{\beta_2}+\frac{\bar{\beta_3}}{2}\right)\right\}^{-1} \ ,\nn
\eea
and
\bea
c(x)&=&1+(\bar{\gamma_1}-\bar{\beta_1}\bar{\gamma_0})x+\frac{1}{2}\left((\bar{\gamma_1}-\bar{\beta_1}\bar{\gamma_0})^2+\bar{\gamma_2}+\bar{\beta_1}^2\bar{\gamma_0}-\bar{\beta_1}\bar{\gamma_1}-\bar{\beta_2}\bar{\gamma_0}\right)x^2 \\
&&+\left(\frac{1}{6}(\bar{\gamma_1}-\bar{\beta_1}\bar{\gamma_0})^3+\frac{1}{2}(\bar{\gamma_1}-\bar{\beta_1}\bar{\gamma_0})(\bar{\gamma_2}+\bar{\beta_1}^2\bar{\gamma_0}-\bar{\beta_1}\bar{\gamma_1}-\bar{\beta_2}\bar{\gamma_0})\right.\nn\\
&&\left.+\frac{1}{3}(\bar{\gamma_3}-\bar{\beta_1}^3\bar{\gamma_0}+2\bar{\beta_1}\bar{\beta_2}\bar{\gamma_0}-\bar{\beta_3}\bar{\gamma_0}+\bar{\beta_1}^2\bar{\gamma_1}-\bar{\beta_2}\bar{\gamma_1}-\bar{\beta_1}\bar{\gamma_2})\right)x^3 \ , \nn
\eea
with $\bar{\gamma_i}=\gamma_i/\beta_0$ and $\bar{\beta_i}=\beta_i/\beta_0$.
The values of the various constants are, in the large $N_c$ limit,
\bea
r_1&=&\frac{17N_c}{8}\\
r_2&=&\frac{N_c^2}{768}(7431-160\pi^2-1920\zeta(3))\nn\\
r_3&=&\frac{N_c^3}{497664}\left(25999999-1095264\pi^2-11200032\zeta(3)+1607040\zeta(5)\right)\nn
\eea
\bea
\beta_0&=\frac{11N_c}{12}\ ,\quad
\beta_1&=\frac{17N_c^2}{24}\ ,\quad\quad
\beta_2=\frac{2857N_c^3}{3456}\ ,\quad\quad
\beta_3=\frac{N_c^4\left(150653-2376\zeta(3)\right)}{124416} \nn\\
\gamma_0&=\frac{3N_c}{8}\ , \quad
\gamma_1&=-\frac{185N_c^2}{384}\ , \quad
\gamma_2=\frac{11413N_c^3}{13824} \  , \quad
\gamma_3=\frac{N_c^4(460151+74048\zeta(3)-126720\zeta(5))}{294912} \ . \nn
\eea
The $\overline {\rm MS}$ mass anomalous dimensions are taken from Ref. \cite{Tarrach:1980up,Chetyrkin:1997dh}.

Finally, we remind that, strictly speaking, we can only fix the ratio between the 
decay constant and the derivative of the mass. We have fixed 
this ambiguity by arbitrarily imposing the $n$ dependence of the 
mass spectrum. 

\subsection{NLO matching}

We now want to obtain extra information on the decay constant by demanding the 
validity of the OPE at ${\cal O}(1/Q^2)$, in particular the absence 
of condensates at this order. We insert the NLO expressions
for $M_X^2(n)$ and $F_X^2(n;\mu)$ into Eq. (\ref{BEuler}) and impose that there be no $1/Q^2$ contribution. 
With the ansatz we are using for the mass at NLO, it is compulsory to introduce the (logarithmically modulated) 
$1/n$ corrections to the decay constant if we want this constraint to hold. Note that it is possible 
to shift all the perturbative corrections to the decay constant.

Imposing that the $1/Q^2$ term vanishes produces the following sum rule:
\bea
\label{matchingNLO}
&&
A_X{d \over dQ^{2}}{\cal B}^{pt.}_X-{2A_X \over Q^2}{\cal B}^{pt.}_X
+{1 \over Q^2}\left[\sum_{n=0}^{n^*-1}F_X^2(n;\mu)
- \int_0^{n^*}dn n F_{X,LO}^2(n;\mu) \right]
+
{F_X^2(n^*;\mu) \over 2Q^2} \nn
\\
\nn
&&
\left.
+
{1 \over Q^2}
\sum_{k=1}^{\infty}(-1)^k{|B_{2k}| \over (2k)!}{d^{(2k-1)} \over dn^{(2k-1)}}
F_X^2(n;\mu) \right|_{n=n^*}
-Q^4 \int_0^{n^*}dn{ F_{X,1}^2(n;\mu) \over (Q^2+M_{X,LO}^2(n))^3}
\\
&&
+Q^4 \int_0^{\infty}dn{F_{X,1}^2(n;\mu)\over (Q^2+M_{X,LO}^2(n))^3}=0
\,.
\eea
This equality should hold independently of the value of $n^*$, which 
formally should be taken large enough so that $\als(n^*B_V) \ll 1$, 
i.e. the limit $\Lambda_{\MS} \ll n^*B_V \ll Q^2$.
Only a few terms in Eq. 
  (\ref{matchingNLO}) can generate $\ln Q^2$ terms, which should cancel at any order. 
Those are the first two and the last two terms. Actually, the next to last term does 
not generate logarithms, but it allows to regulate possible infrared divergences appearing 
in the calculation.  
Therefore, asking for the cancellation of the $1/Q^2$ suppressed logarithmic terms 
produced by the first two and the last term in Eq. (\ref{matchingNLO}) 
fixes  $F^2_{X,1}$. The non-logarithmic terms should also 
be cancelled but they cannot be fixed from perturbation theory. 

One can actually find an explicit solution to the above constraint for $F^2_{V,1}$ by performing 
some integration by parts. We obtain 
\be
F^2_{X,1}(n;\mu)=\frac{A_{X}}{B_{X}}\frac{d}{dn}\left(n F^2_{X,0}(n;\mu)\right) \ .
\ee
Note that $F^2_{X,1}(n;\mu)$ is of the same order in $\als$ as $F^2_{X,0}(n;\mu)$. 
This is different that in the vector/axial-vector case. For further details to the procedure we refer to Ref. \cite{Mondejar:2007dz}. 
Here we would only like to  insist that non-logarithmic terms should also be cancelled, 
but they cannot be fixed from perturbation theory. For these terms we cannot even give a closed expression: $\ln Q^2$-independent 
terms may receive contributions 
from any subleading order in the $1/n$ expansion of the masses and decay constants. The reason is
 that the decay constant at a given order in $1/n$ is obtained after performing some integration by 
parts, which generates new ($\ln Q^2$-independent) terms that can be $Q^2$ enhanced.
This statement is general and also applies to any subleading power in the $1/Q^2$ matching 
computation.

\subsection{NNLO matching}

We now consider expressions for the mass and decay constants at NNLO.
For the first time we have to consider condensates. Simplifying terms that do not produce logs, 
we obtain the following equation,
\bea
&&
-\frac{1}{N_cQ^2}\frac{3}{22}\lp \frac{\als(Q^2)}{\als(\mu^2)}\rp^{2\bar{\gamma_0}}\beta(\als(\mu))\langle vac |G^2(\mu)| vac \rangle \\
&&\doteq Q^4\int_{n^*}^{\infty}\frac{dn}{(Q^2+B_X n)^3}\lp \frac{F_{X,2}^2(n)}{n}-\frac{1}{B_{X}}\frac{d}{dn}
\left(\frac{1}{2}A_{X}F^2_{X,1}(n;\mu)+C_{X}F^2_{X,0}(n;\mu)\right)\right) \ , \nn
\eea
where $\doteq$ stands for the fact that this equality is only true at leading logarithmic order. Using
\be
\frac{1}{Q^2}\lp \frac{\als(Q^2)}{\als(\mu^2)}\rp^{2\bar{\gamma_0}} \doteq -Q^4\int_{n^*}^{\infty} dn\frac{1}{(Q^2+B_X n)^3}\frac{1}{n}2\bar{\gamma_0} \lp\frac{\als(B_X n)}{\als(\mu^2)}\rp^{\bar{2\gamma_0}}\frac{\beta_0}{\pi}\als(B_X n)  \ ,
\ee 
we have
\bea
F^2_{X,2}(n;\mu)&=&\frac{n}{B_{X}}\frac{d}{dn}\left(\frac{1}{2}A_{X}F^2_{X,1}(n;\mu)+C_{X}F^2_{X,0}(n;\mu)\right) \\
&&+\frac{3}{12\pi}\bar{\gamma_0} \lp\frac{\als(B_X n)}{\als(\mu^2)}\rp^{2\bar{\gamma_0}}N_c\als(B_X n)\frac{\beta(\als(\mu))\langle vac |G^2(\mu)| vac \rangle}{N_c^2} \ . \nn
\eea
Note that the accuracy of this result is set 
by our knowledge of the matching coefficient of the gluon condensate. Note as well 
that $F^2_{X,2}(n;\mu)$ is $\als$ suppressed with respect to $F^2_{X,1}(n;\mu)$ and 
$F^2_{X,0}(n;\mu)$.

\subsection{Scalar versus pseudoscalar correlators}
From the previous analysis it is evident that the coefficients 
of the mass spectrum are free parameters and can be different for the scalar and 
pseudoscalar channel, in other words, they cannot be fixed from the OPE alone. 
This point was already emphasized in Ref. \cite{Cata:2006fu} for a model that reproduces 
the parton-model logarithm. In this paper, we show that that the inclusion of corrections in 
$\als$ does not affect that conclusion, and that $B_S$ and $B_P$ somewhat play 
the role of the renormalization scale in the analogous perturbative analysis in the Euclidean 
regime, and are therefore unobservable. Overall, the situation is similar to the 
case of vector and axial-vector correlators studied 
in \cite{Golterman:2002mi, Mondejar:2007dz}.

However, although the constants that characterize the spectrum can be different for the scalar and 
pseudoscalar channels, they have to produce the same expressions for the OPE when combined with the 
decay constants. This produces some relations that must be fulfilled. Defining $t\equiv B_S n=B_P n'$ 
and taking $n$ and $n'$ as continuous variables, these relations are
\be
\frac{1}{B_X^2}F_{X,0}^2(n;\mu)= \frac{1}{t}\frac{1}{\pi}\text{Im} \Pi_X^{pert.}(t;\mu) \equiv \frac{1}{t}f_0(t;\mu) \ ,
\ee
\be
\frac{1}{A_X B_X}F_{X,1}^2(n;\mu)=\frac{d}{dt}f_0(t;\mu) \ ,
\ee
\bea
&&
F_{X,2}^2(n;\mu)-\frac{n}{B_{X}}\frac{d}{dn}\left(\frac{1}{2}
A_{X}F^2_{X,1}(n;\mu)+C_{X}F^2_{X,0}(n;\mu)\right)
\\
\nn
&&
= \frac{3}{12\pi}\bar{\gamma_0} \lp\frac{\als(t)}{\als(\mu^2)}\rp^{\bar{2\gamma_0}}N_c\als(t)\frac{\beta(\als(\mu))\langle vac |G^2(\mu)| vac \rangle}{N_c^2}\ .
\eea

\subsection{Numerical analysis}
The aim of this section is not to perform an in-depth numerical analysis of the expressions we have found.
We lack experimental data for resonances at high excitations, where our expansion in $1/n$ would work best, 
and have no information on the decay constants; also, one must not forget that we are staying at leading order in the $1/N_c$ expansion 
and in the chiral limit. Instead, the purpose of our analysis is just to get a feeling of the relative size of the corrections in $1/n$, 
and of the importance of the resummation of the powers of $\als$ in (\ref{LO}).
We restrict ourselves to the SU(2) flavour case. In the table we show, in parenthesis, the experimental values of the masses of the mesons. 
We fit these values to Eq. (\ref{massn}), and from these fits we find the decay constants. We take most of the experimental values from \cite{PDG}, 
except the pseudoscalar states with $n=3,4$, which we take from \cite{CrystalBarrel}. We do not include the pion (the state with $M^2=0$) as input 
in our fit. We define
\be
\label{G}
F^2_{S,P}(n;\mu)\equiv 2 B_{S,P}n G^2_{S,P}(n;\mu) \ .
\ee

\begin{table}[h!!!]
\addtolength{\arraycolsep}{0.2cm}
$$
\begin{array}{|l||c|c|c|c|}
\hline
  & n=1 & n=2 & n=3 & n=4   
\\ \hline\hline
M_{f_0} &  986 (980\pm 10) & 1342 (1370) & 1544 (1507\pm 5)  & 1703 (1718\pm 6)  
\\ \hline\hline
M_{\pi} & 1305 (1300\pm 100)  & 1791 (1812\pm 14)  & 2098 (2070\pm 35) & 2349 (2360\pm 30)  
\\ \hline\hline
G_{S} & 13318 & 452  & 229 & 188
\\ \hline\hline
G_{P} & 649 & 301 & 255 & 233 
\\  \hline
\end{array}
$$
\caption{{\it We give the experimental values of the masses (in MeV) for scalar and pseudoscalar particles (within parenthesis), compared with 
the values obtained from the fit. 
We take $\als(1\, {\rm GeV})=0.5$ and 
$\beta\langle G^2 \rangle=-(352 \,{\rm MeV})^4$.
Note that the values of $G_S$ and $G_P$ depend on the factorization scale. We have taken 
$\mu^2=10B_S$ and $\mu^2=10B_P$ for the scalars and pseudoscalars respectively.}}
\label{table}
\end{table}

The parameters of the mass spectrum obtained from the fit to the experimental values in the table are

\bea
B_S=0.456 \, {\rm GeV}^2
\qquad
A_S&=&1.262 \, {\rm GeV}^2
\qquad
C_S=-0.746\, {\rm GeV}^2
\\
B_P=1.040\, {\rm GeV}^2
\qquad
A_P&=&1.589\, {\rm GeV}^2
\qquad
C_P=-0.926\, {\rm GeV}^2
\,.
\eea

\begin{figure}[hhh]
\centering
\includegraphics[width=0.485\columnwidth]{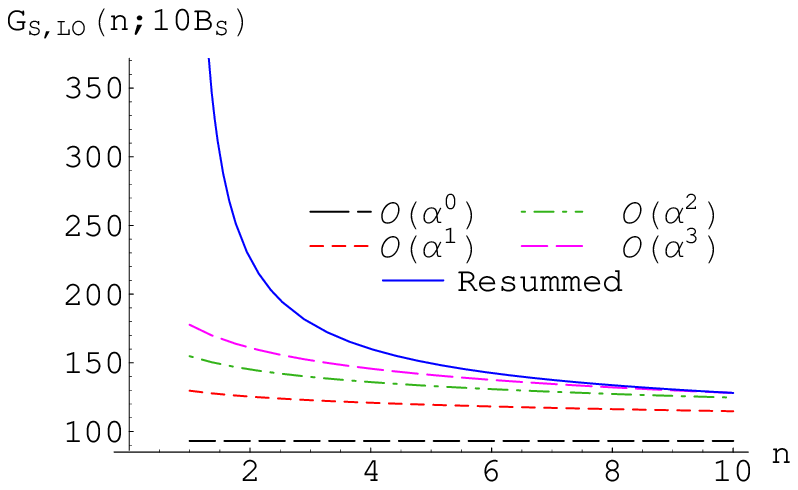}
\hspace{0.05 in}
~\includegraphics[width=0.485\columnwidth]{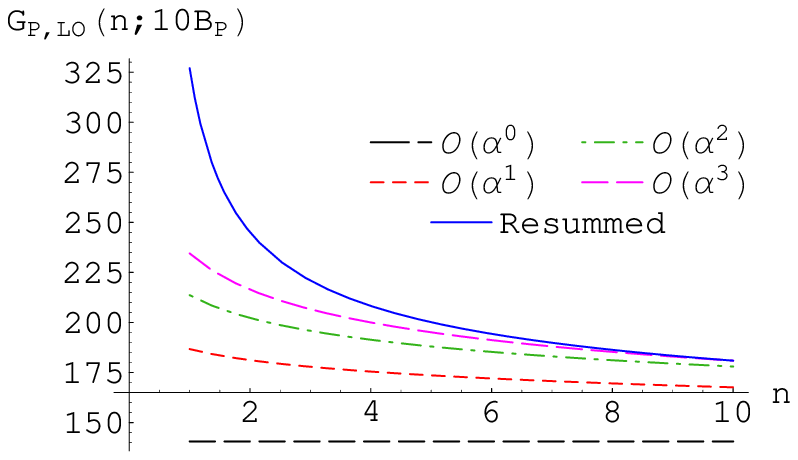}
\caption{\it In this plot we show the differences between the expanded and the resummed expression for $G_{(S,P),LO}(n;\mu)$, following the definition given in eq. (\ref{G}).}
\label{resum}
\end{figure}

In Figure \ref{resum} we can see the difference between the resummed expression 
of $G_{X,LO}^2(n;\mu)$, as taken from Eq. (\ref{LO}), and the expanded expressions, 
at different orders in $\als$. We can see important differences with respect to the analysis 
of Ref. \cite{Mondejar:2007dz} for the vector and axial-vector currents. 
The impact of the resummation of logarithms (or of perturbation 
theory in general) is much more important for the scalar and pseudoscalar currents, in 
particular for the former. Note as well that the importance of these perturbative corrections 
enforces us to take a large value for the factorization scale to ensure nice convergence 
properties of the perturbative series. 
By inspection of the plots we conclude that our figures are a reliable 
approximation for, at least, $n=2$ (being on the boundary) or larger. 
This should be compared with the vector and axial-vector case, in which even for $n=1$ 
one obtains reasonable numbers. One should keep in mind, however, 
that the decay constants we are computing here are scheme and scale dependent quantities. 
Therefore, it would be more appropriated to consider them in combination with another quantity with the 
inverse scheme and scale dependence to become a direct observable quantity. 

\begin{figure}[h!!!]
\centering
\includegraphics[width=0.485\columnwidth]{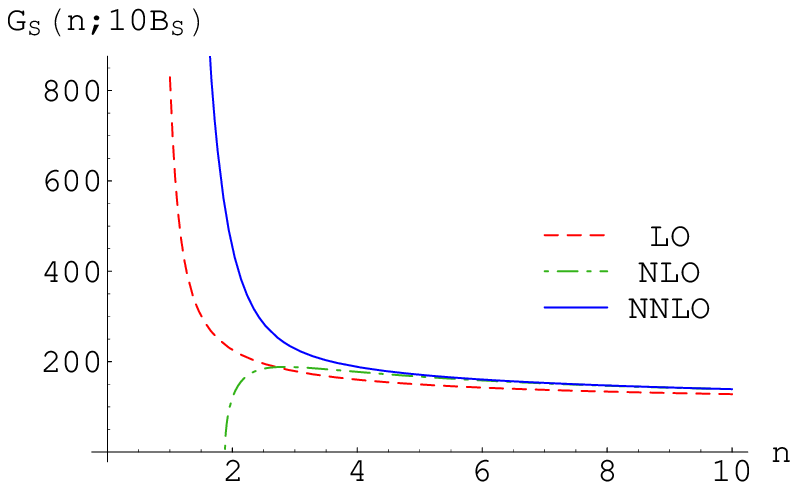}
\hspace{0.02 in}
~\includegraphics[width=0.485\columnwidth]{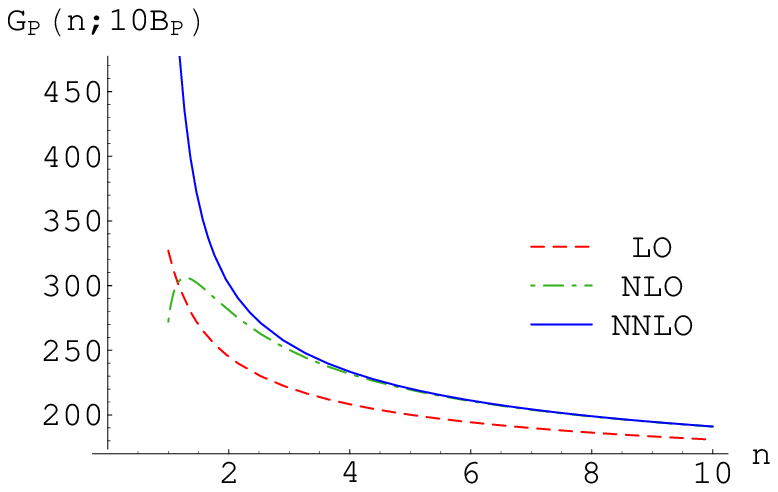}
\caption{\it In this plot we show $G_{S}(n;\mu)$ and $G_{P}(n;\mu)$ 
at different orders in the $1/n$ expansion.}
\label{1/n}
\end{figure}

Figure \ref{1/n} shows us that the dependence of the decay constants on $n$ is small (again 
from $n \sim 2$ on). The corrections in $1/n$ are rapidly converging, 
the result at NNLO being almost indistinguishable from that of NLO in the region where the 
series is convergent. One may ask whether these results change qualitatively by choosing 
a different value of the gluon condensate. We remind that the value for the gluon condensate 
will vary depending on the renormalon subtraction scheme used. We have seen that the 
change is quite small if we vary the gluon condensate between the range 0.01 and 0.04 GeV$^4$.

\section{Preasymptotic effects in $1/n$ in the 't Hooft Model}

A systematic analysis of the preasymptotic effects in $1/n$ has not been 
undertaken until the analysis of Ref. \cite{Mondejar:2007dz} and this paper. 
Therefore, it is interesting 
to check the methodology in a specific model where the results obtained with the 
OPE can be tested with the exact results. Here we consider the 't Hooft model, 
which will allow us to perform such an analysis. In this model, the preasymptotic 
corrections to the mass spectrum are known \cite{hooft2} (actually they have a 
$\ln n$ dependence, which will prove to be crucial) but not those to 
the decay constants, for which the analysis of the preasymptotic 
effects in $1/n$ is not an easy issue.

Unlike in the rest of the paper, we will consider the general situation with 
non-zero (and different) quark masses. We will focus our interest in the 
scalar and pseudoscalar currents. One reason is that in two dimensions the 
vector and axial-vector current matrix elements can be related to them. 
Nevertheless, the main motivation is that 
we are interested in situations where logarithms of $Q$ can be generated, which can only 
happen if an infinite number of resonances contributes to the correlator. This 
does not happen for the vector and axial-vector correlator in the massless limit, where 
they both become almost trivial, since only one resonance (the ground state) has overlap 
with the current\footnote{In fact, the combination of these results 
with the behaviour of the ground state wave function in the $x \rightarrow 0$ limit (or 
in other words its mass) is used to fix the quark condensate \cite{Zhitnitsky:1985um}.}\cite{callan,einhorn}.

\subsection{Preasymptotic effects in $1/n$ from the OPE}
\label{OPE2D}

The main aim here is to repeat up to NLO and in two dimensions 
the analysis performed in sec. \ref{sec4D}.  
The difference is that now we know the spectrum at NLO, which reads \cite{hooft2}
\be
\label{massn2D}
M_{X}^2(n)=M^2(n)=\sum_{s=-1}^{\infty}B^{(-s)}n^{(-s)}=B^{(1)}n+B^{(0)}+\cdots
\,,
\ee
and we define $M^2_{LO}(n)=B^{(1)}n$.  
Note that the spectrum is the same for the scalar and pseudoscalar channel. 
The coefficients $B^{(1)}$ and $B^{(0)}$ are known in the 't Hooft model. 
They read \cite{hooft2}
\be
B^{(1)}=\pi^2 \beta^2 \,,\qquad B^{(0)}=(m_{i,R}^2+m_{j,R}^2)\ln n+{\rm constant}
\,,
\ee
where $i$ and $j$ represent, respectively, the flavor of the quark and antiquark that make up the meson in the 
$\text{'t Hooft}$ model, $\beta^2 =g^2N_c/(2\pi)$, and $m_{i,R}^2= m_i^2-\beta^2$ 
is the renormalized mass. The explicit expression for the constant term  can be 
found in Ref. \cite{Brower:1978wm}. 

Similarly to the mass we write
\be
\label{F}
F_X^2(n)=F_{X,0}^2(n) + \frac{F_{X,1}^2(n)}{n}+\dots
\,,
\ee
and $F_{X,LO}^2(n)=F_{X,0}^2(n)$, $F_{X,NLO}^2(n)=F_{X,0}^2(n)+F_{X,1}^2(n)/n$. 
Note that, unlike our former expression for the decay constants, Eq. (\ref{decayn}), 
there is no $n$ multiplying in front. This is due to the fact that in two dimensions the correlator $\Pi_X$ goes like
\be
\Pi_{X}(q) \sim \ln{\frac{Q^2}{\mu^2}} \ .
\ee
The correlator has also no anomalous dimension in the 't Hooft model. 
The definition of the correlators used in this section is the same to the one 
given in Eq. (\ref{correlatorSP}) but with general currents (with in principle different 
flavours): $J_S=\bar \psi_j\psi_i$ and $J_P=i\bar \psi_j\gamma_5\psi_i$. 
The hadronic expressions for the correlators read 
\bea
\Pi^{hadr.}_{S}&=&\sum_{n=1,3,5\dots}^{\infty}\frac{F_S^2(n)}{M^2(n)-q^2-i\epsilon} \,,
\\
\Pi^{hadr.}_{P}&=&\sum_{n=0,2,4\dots}^{\infty}\frac{F_P^2(n)}{M^2(n)-q^2-i\epsilon} \ ,
\eea
where $F_X(n)=\langle vac |J_X|ij;n\rangle$. 
We note that only odd and even states contribute to the scalar and pseudoscalar 
correlator respectively (this implies that for the scalar correlator the ground state does not 
contribute). This result was obtained in Ref. \cite{callan}, using some general properties 
of the 't Hooft equation operator. Those properties can be understood somewhat on a 
pure symmetry basis\footnote{We cannot 
use parity symmetry due to the fact that the quantization frame that we use 
does not respect this symmetry explicitly. This means that the states and 
currents have complicated transformation properties under this discrete symmetry.}.  
Nevertheless, it is not an easy task to visualize 
this symmetry at the Lagrangian level. The situation is similar to the one 
found in the spectrum of diatomic atoms \cite{Messiah:1979eg} or of NRQCD 
in the static limit (see for instance \cite{Brambilla:2004jw}). 
In those situations a convenient way to deal with the problem is to project 
the Lagrangian to the Hilbert space sector we are interested in. In our case, this is 
basically equivalent to ending up in the 't Hooft equation (wich we present in the next 
section), where one can use the operator identities found in Ref. \cite{callan} for the current 
matrix elements (however it should also be possible to relate these operator identities 
with some underlying symmetries of the system), in particular Eq. (\ref{sym}).
Using these results one can discriminate which states give non-zero contribution to 
each correlator.

We now define the Adler-like function in two dimensions
\be
\mathcal{A}_X\equiv -Q^2\frac{d\Pi_X(Q^2)}{dQ^2} \ ,
\ee
for which its hadronic expressions read
\be
\mathcal{A}^{hadr.}_{X}=Q^2\sum_{n_X}^{\infty}\frac{F_X^2(n)}{(M(n)^2+Q^2)^2} \ ,
\ee
where $n_X$ stands for the sum for the scalar or pseudoscalar case respectively.

\begin{figure}[htb]
\centering
\includegraphics[width=0.19\columnwidth]{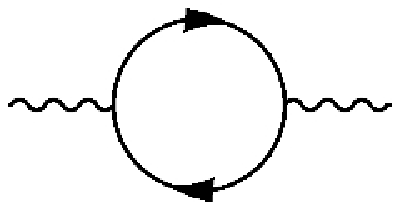}
\hspace{24in}
\vspace{0.5cm}
~\includegraphics[width=0.65\columnwidth]{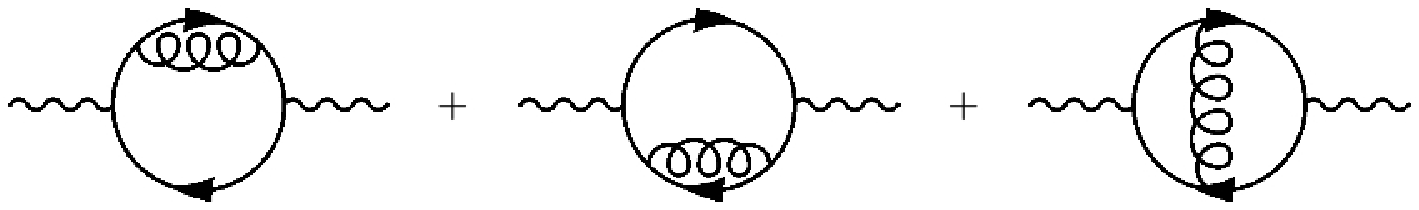}
\caption{\it Diagrams that contribute to the OPE at one loop in 
perturbation theory.}
\label{diagrams}
\end{figure}

In order to follow the procedure used in the previous section we need the OPE 
expression for $\mathcal{A}_X$. It reads:
\bea
&&
\nn
\mathcal{A}^{pert.}_{X}=\frac{N_c}{2\pi}
\left(1+\frac{\beta^2}{Q^2}
+D_X\frac{m_im_j}{Q^2}\left(\text{ln}\lp \frac{m_i^2}{Q^2}\rp
+\text{ln}\lp \frac{m_j^2}{Q^2}\rp\right) 
+\frac{2\pi}{N_c} 
D_X\frac{m_j\langle\bar \psi_i \psi_i \rangle+m_i\langle\bar \psi_j \psi_j \rangle}{Q^2}
\right) 
\\
&&
+
\cdots\ ,
\label{AXpert}
\eea
where $D_S=1$ and $D_P=-1$, and we have neglected terms of ${\cal O}(m^2/Q^2)$ 
(except for the logarithm term) and terms of ${\cal O}(1/Q^3)$.
The pure perturbative piece of this result is obtained from the computation of the 
diagrams shown in Fig. \ref{diagrams}. The condensate contribution is obtained by 
``collapsing" one of the quark propagators. 
We see that the coefficient of the quark condensate has no $\ln Q$ piece. 

Note that in two dimensions the coupling constant $\beta^2$ has dimensions. 
Therefore, the perturbative and OPE expansions are related. In particular, 
in two dimensions, there are no problems with renormalons, due to the fact that 
there are no marginal operators in the Lagrangian.

In Eq. (\ref{AXpert}) the quark condensate, as usual, only includes the 
pure non-perturbative contribution, ie. the free result\footnote{Note that 
we have a sign difference with respect to the result of Ref. \cite{Burkardt:1995eb}. We also 
disregard finite pieces, which are scheme dependent.} 
\be
\langle\bar \psi_i \psi_i \rangle_{pert.}=\frac{N_c}{2\pi}m_i\ln{\frac{m_i^2}{\mu^2}}
\ee
has been subtracted and displayed explicitly in  $\mathcal{A}^{pert.}_{X}$.

We can now combine all this information with our knowledge of the 
large $n$ behaviour of the mass spectrum to obtain the $1/n$ 
preasymptotic effects of the decay constants.

We follow the same procedure of sec. \ref{matching}. To rewrite $\mathcal{A}^{hadr.}_{X}$ 
using the Euler-Maclaurin formula, we first define $n=2n'+1$ for the scalar case and $n=2n'$ 
for the pseudoscalar case, so that we can transform the sums into an integral over a continuous variable. 
The only term that will give a LO contribution, not suppressed by powers of $Q^2$, 
will be
\be
Q^2\int_0^{\infty}\frac{dn}{2}\frac{F_{X,LO}^2(n)}{(M_{LO}^2(n)+Q^2)^2}
=\int_0^{\infty}\frac{dn}{2}\frac{F_{X,0}^2(n)}{(n\pi^2\beta^2+Q^2)^2}=\frac{N_c}{2\pi} \ ,
\ee
where the last equality is the matching to the partonic result and gives us
\be
\label{F02D}
F_{X,0}^2(n)=N_c\pi\beta^2 \ .
\ee

In order to go beyond the parton result we need to consider 
the $1/n$ corrections to the linear behaviour in the mass.  
Note that the form of the spectrum is different from the one assumed in Ref. 
\cite{Mondejar:2007dz}, or in sec. \ref{sec4D}: in the 't 
Hooft model we know that $B^{(0)}$ is not a constant and diverges logarithmically 
in $n$. We actually do not care about the finite piece of $B^{(0)}$, since such 
contribution would constrain the ${\cal O}(1/n^2)$ term of the decay constant 
but not the ${\cal O}(1/n)$ term; this is so because $F^2_{X,0}$ is constant. Therefore in what 
follows we will neglect the constant term in $B^{(0)}$ and only consider 
the logarithmic correction. 
Within this approximation, the matching at NLO reads (at logarithmic order, after 
expanding and integrating by parts),
\bea
&&Q^2\int_{n^*}^{\infty}\frac{dn}{2}\frac{F_{X,1}^2(n)/n}{(n\pi^2\beta^2+Q^2)^2}
- Q^2\int_{n^*}^{\infty}\frac{dn}{2}\frac{F_{X,0}^2}{(n\pi^2\beta^2+Q^2)^2}
\frac{(m_{i,R}^2+m_{j,R}^2)}{n \pi^2 \beta^2}\\
&&\doteq -\frac{N_c}{2\pi}\frac{D_X}{Q^2}\lp 2m_im_j\rp\text{ln}(Q^2)\nn
\doteq -Q^2 \int_{n^*}^{\infty}\frac{dn}{2}\frac{1}{(n\pi^2\beta^2+Q^2)^2}N_c 
\pi \beta^2 \frac{D_X(2m_im_j)}{n\pi^2\beta^2} \ , 
\eea
where again $\doteq$ means that the equality is only true at logarithmic order. 
And so,
\be
\label{F12D}
\frac{F_{X,1}^2(n)}{n}=-N_c\pi \beta^2\frac{2m_im_jD_X-m_{i,R}^2-m_{j,R}^2}{n\pi^2\beta^2} \ .
\ee

It will prove useful in the next section to combine the scalar and pseudoscalar 
result for the decay constant in a single function. We define
\begin{eqnarray*}
F^2(n)=\,\left\{
\begin{array}{ll}
&
F_S(n) \, {\rm for} \, n=\, {\rm odd}
\\
&
F_P(n) \, {\rm for} \, n=\, {\rm even}\,.
\end{array}  
\right.
\end{eqnarray*}
Therefore, we have
\be
\label{F2pert}
F^2(n)=N_c\pi \beta^2
\left[
1+\frac{m_{i,R}^2+m_{j,R}^2}{n\pi^2\beta^2}
+\frac{2m_im_j}{n\pi^2\beta^2}(-1)^n
\right]
\,.
\ee
Let us note that there is a term which is sign-alternating. 

Finally, we stress that we have obtained these expressions for the decay constant using the OPE, 
symmetries 
and the knowledge of the mass spectrum, nothing else. In the next subsection we check whether 
the explicit numeric computation of the decay constants from the 't Hooft equation 
fulfills these expectations.

\subsection{Preasymptotic effects in $1/n$ from the hadronic solution}

In the 't Hooft model it is possible to write the decay constants $F_X^2(n)$ in terms 
of the light-cone distribution amplitude of the bound state, $\phi^{ij}_n(x)$, 
which is the solution of the 
equation
\be
\label{thoofteq}
M^2(n)\phi^{ij}_n(x)
= \lp \frac{m^2_{i,R}}{x}+ \frac{m^2_{j,R}}{1-x}\rp
\phi^{ij}_n(x)-\beta^2\int_0^1dy \phi^{ij}_n(y)\mathrm{P}\frac{1}{(y-x)^2}
\,,
\ee
where $x=p^+/P_n^+$, with $p^+$ 
being the momentum of the quark $i$, and P stands for Cauchy's Principal Value. 

The decay constant for the scalar case then reads \cite{callan}
\be
\label{Fshad}
F_S(n)=\sqrt{\frac{N_c}{4\pi}}\int_0^1dx\phi^{ij}_n(x)\left(\frac{m_i}{x}-\frac{m_j}{1-x}\right)
=
m_i\sqrt{\frac{N_c}{\pi}}\int_0^1dx\frac{\phi^{ij}_n(x)}{x} \;\; \text{for $n$ odd}\,,
\ee
and zero otherwise. 
For the pseudoscalar 
we have
\be
\label{Fphad}
F_P(n)=\sqrt{\frac{N_c}{4\pi}}\int_0^1dx\phi^{ij}_n(x)\left(\frac{m_i}{x}+\frac{m_j}{1-x}\right)
=
m_i\sqrt{\frac{N_c}{\pi}}\int_0^1dx\frac{\phi^{ij}_n(x)}{x} \;\; \text{for $n$ even}
\,,
\ee
and zero otherwise.
Note that in the above two equalities we have used the remarkable identity
\be
\label{sym}
m_i\int_0^1dx\frac{\phi^{ij}_n(x)}{x}=(-1)^nm_j\int_0^1dx\frac{\phi^{ij}_n(x)}{1-x}
\,,
\ee 
obtained in Ref. \cite{callan}. 

By comparing Eqs. (\ref{Fshad}) and (\ref{Fphad}) with Eq. (\ref{F2pert}) we can 
obtain expressions for $\int_0^1dx\frac{\phi^{ij}_n(x)}{x}$ with $1/n$ 
accuracy\footnote{
In this result the global phase has been fixed to 1. One always has this freedom but note 
that this also fixes the global phase of $\phi_n^{ij}(x)$, which is no longer arbitrary. 
The leading order result was first obtained in 
Refs. \cite{callan,einhorn} and later confirmed using the 
boundary layer equation in Ref. \cite{Brower:1978wm}. The $1/n$ correction is new. 
Note that these $1/n$ corrections have to be included in any analysis of 
preasymptotic effects in the 't Hooft model, in particular in the analysis 
of Ref. \cite{Mondejar:2006ct}. Nevertheless, the results for the moments obtained in that 
paper remain unchanged.}: 
\be
\label{int1}
\int_0^1dx\frac{\phi^{ij}_n(x)}{x}=\pi \frac{\beta}{m_i}
\left[
1+\frac{m_{i,R}^2+m_{j,R}^2}{2n\pi^2\beta^2}+\frac{m_im_j}{n\pi^2\beta^2}(-1)^n
+{\cal O}\left(\frac{1}{n^2}\right)
\right]
\,.
\ee
By using symmetries and the 't Hooft equation, we can also obtain 
\bea
\label{int0}
M^2(n)\int_0^1dx\phi_n^{ij}(x)
&=&
m_i^2\int_0^1dx \frac{\phi^{ij}_n(x)}{x}+m_j^2\int_0^1dx \frac{\phi^{ij}_n(x)}{1-x}\\
\nn
&=&\pi {\beta}{m_i}
\left[
1+\frac{m_{i,R}^2+m_{j,R}^2}{2n\pi^2\beta^2}+\frac{m_im_j}{n\pi^2\beta^2}(-1)^n
\right]
\\
&&
+
(-1)^n
\pi {\beta}{m_j}
\left[
1+\frac{m_{i,R}^2+m_{j,R}^2}{2n\pi^2\beta^2}+\frac{m_im_j}{n\pi^2\beta^2}(-1)^n
\right]
+{\cal O}\left(\frac{1}{n^2}\right)
\,,
\nn
\eea
where in the last equality we have made use of the symmetry property of the 't Hooft wave functions,
\be
\phi^{ij}_n(x)=(-1)^n\phi^{ji}_n(1-x) \ .
\ee

These results can also be used to deepen our analytic understanding of the 
't Hooft wave function in the end-point regions in the massless limit 
(actually we need only one mass to go to zero at each boundary, 
$m_i$ for $x\to 0$, or $m_j$ for $x\to 1$). 
In the massless quark limit the integral in Eq. (\ref{int1}) 
is dominated by the behaviour of the wave-function in the 
$x \rightarrow 0$ boundary:
\be
\phi^{ij}_n(x)=c_nx^{\beta_i}\left(1+o(x)\right) \ ,
\ee
where $\beta_i$ is the solution of 
\be
m_{i,R}^2+\beta^2\pi\beta_i\cot{\pi\beta_i}=0 \ ,
\ee
which in the massless limit approximates to
\be
\beta_i=\frac{\sqrt{3}}{\pi}\frac{m_i}{\beta}+o(m_i)
\,.
\ee
Therefore we obtain for the integral
\be
\label{cnFn}
\lim_{m_i\rightarrow 0} 
m_i\int_0^1dx\frac{\phi^{ij}_n(x)}{x}=c_n\frac{\pi \beta}{\sqrt{3}}  \,,
\ee
and, in the massless quark limit, the problem of getting the 
decay constants could be reformulated as that of obtaining the 
coefficient $c_n$.
One can find its value for $n=0$ (note that this result does also require $m_j=0$):
\be
\label{c0}
\lim_{m_{i,j}\rightarrow 0}c_0=1 .
\ee 
In principle, there are several ways to 
obtain this result. One can work along the lines of Ref. \cite{hooftproc} 
to obtain an approximated Schrodinger-like equation, which can be approximately solved 
for the ground state. Another possibility to fix the value $c_0$ is by matching the 
solution $\phi_0=1$ (the exact solution in the strict massless limit) and the solution $\phi_0=c_0x^{\beta_i}$ in the region of overlap 
(the latter is valid for $x \ll 1$, 
whereas it can be approximated to a constant, $c_0$, 
for values larger than $e^{-\beta/m_i}$, which is a very small quantity for small 
masses. Therefore, there is a region on which the 
$constant$ solutions: ``$c_0$", and ``1", overlap and should be equal by continuity). 
One can also use the value of $\lim_{m_i\rightarrow 0} \int_0^1\phi_0(x)=1$, to fix $c_0$. 

Nevertheless, in this paper, we are mainly interested in the study of high excitations. 
The combination of Eq. (\ref{cnFn}) and Eq. (\ref{F2pert}) allows us to give the 
following prediction for $c_n$: 
\be
\label{cnasympt}
\lim_{m_i\rightarrow 0}c_n=\sqrt{3}
\left(
1-\frac{1}{\pi^2}\frac{1}{n}
+\frac{m_{j}^2}{2n\pi^2\beta^2}
+{\cal O}(1/n^2)\right)\,.
\ee
How much of all this can be understood by a direct analytic computation? 
In principle the 't Hooft equation can only be solved numerically. Nevertheless, 
for large $n$, the 't Hooft equation can be approximated by the boundary-layer
equation \cite{einhorn}: 
\be
\label{thoofteqlayer}
\phi^i(\xi)= \frac{m^2_{i,R}}{\xi}\phi^i(\xi)
-\beta^2\int_0^\infty d \xi' \phi^i(\xi')
\mathrm{P}\frac{1}{(\xi'-\xi)^2}
\,,
\ee
where 
\be
\phi^i(\xi) \equiv \lim_{n \rightarrow \infty} \phi^{ij}_n(\xi/M^2(n))
\,,
\ee
as far as we are not in the $x \rightarrow 1$ limit (one could also 
write a symmetric equation convenient for $x \sim 1$ region). It is possible 
to analytically solve the Mellin transform of this equation \cite{Brower:1978wm}. 
In particular, one obtains
\be
\label{boundary}
\int_0^{\infty} d\xi \frac{\phi^i(\xi)}{\xi}=\pi\frac{\beta}{m_i}
\,,
\qquad
\int_0^{\infty} d\xi \phi^i(\xi)=\pi \frac{m_i}{\beta}
\ .
\ee
From these results one easily checks the leading order terms of Eqs. (\ref{int1}) 
and Eq. (\ref{cnasympt}). Note that for the last equation $\phi_n(x) \simeq c_nx^{\beta_i}$ 
is valid for $x \ll 1$. In particular, for high
excitations, the region of validity of this expression 
is very small, restricted to the region 
$\xi \equiv M_n^2 x\ll \beta^2$. 

In order to check the $1/n$ corrections in Eqs. (\ref{int1}) and (\ref{cnasympt}), 
one should be 
able to go one order beyond the analysis of Ref. \cite{Brower:1978wm}, which appears to 
be a formidable task. This would go far beyond the aim of this paper. Instead,
in the remainder of this section, we will try to see 
whether Eqs. (\ref{int1}), (\ref{c0}) and (\ref{cnasympt}) can be 
confirmed by a direct numerical computation. 

In order to perform the numerical analysis we will use two methods: 
\begin{enumerate}
\item
One is based on the Brower-Spence-Weis improvement of the Multhopp technique 
\cite{Brower:1978wm}. In this method the 't Hooft wave function is decomposed in 
a basis of sin functions. Therefore, the 't Hooft equation becomes a infinity-dimensional 
matrix, which at the numerical level is truncated  for a finite number of sines. This method 
appears to be very well suited for very high excitations, and allows us to work with 
different quark masses without problems. Nevertheless, it has some 
problems for very low masses and it does not have the right functional behaviour in the $x \rightarrow 
0,1$ limit.
\item
The other method that we use is based on the decomposition of the  't Hooft wave function in 
a basis of $x^{\beta_i}(1-x)^{\beta_j}P(x)$ functions, where $P(x)$ are Jacobi polynomials. 
This method has already been used in Refs. \cite{Bardeen:1979xx,Ditsas:1983ix}. Unfortunately, 
we are only able to give reliable numbers for even states with equal masses. For those states we 
can perform numerical checks with the numbers given in \cite{Ditsas:1983ix} 
(we use them as numerical 
check of our implementation of the method, since our interest is on a different regime than that of \cite{Ditsas:1983ix}: high excitations and 
small masses). In any case, this method does not appear to work very well 
for very high excitations, as well as for very small masses (though it can reach lower limits than method 1). Moreover, at a certain point the introduction of more Jacobi polynomials in the calculation spoils the convergence of the result, except for the wave function of the ground state. On the other hand, 
by construction, this method should have the right functional behaviour in the $x \rightarrow 
0,1$ limit.  
\end{enumerate} 

\begin{figure}[hp]
\centering
\includegraphics[width=0.8\columnwidth]{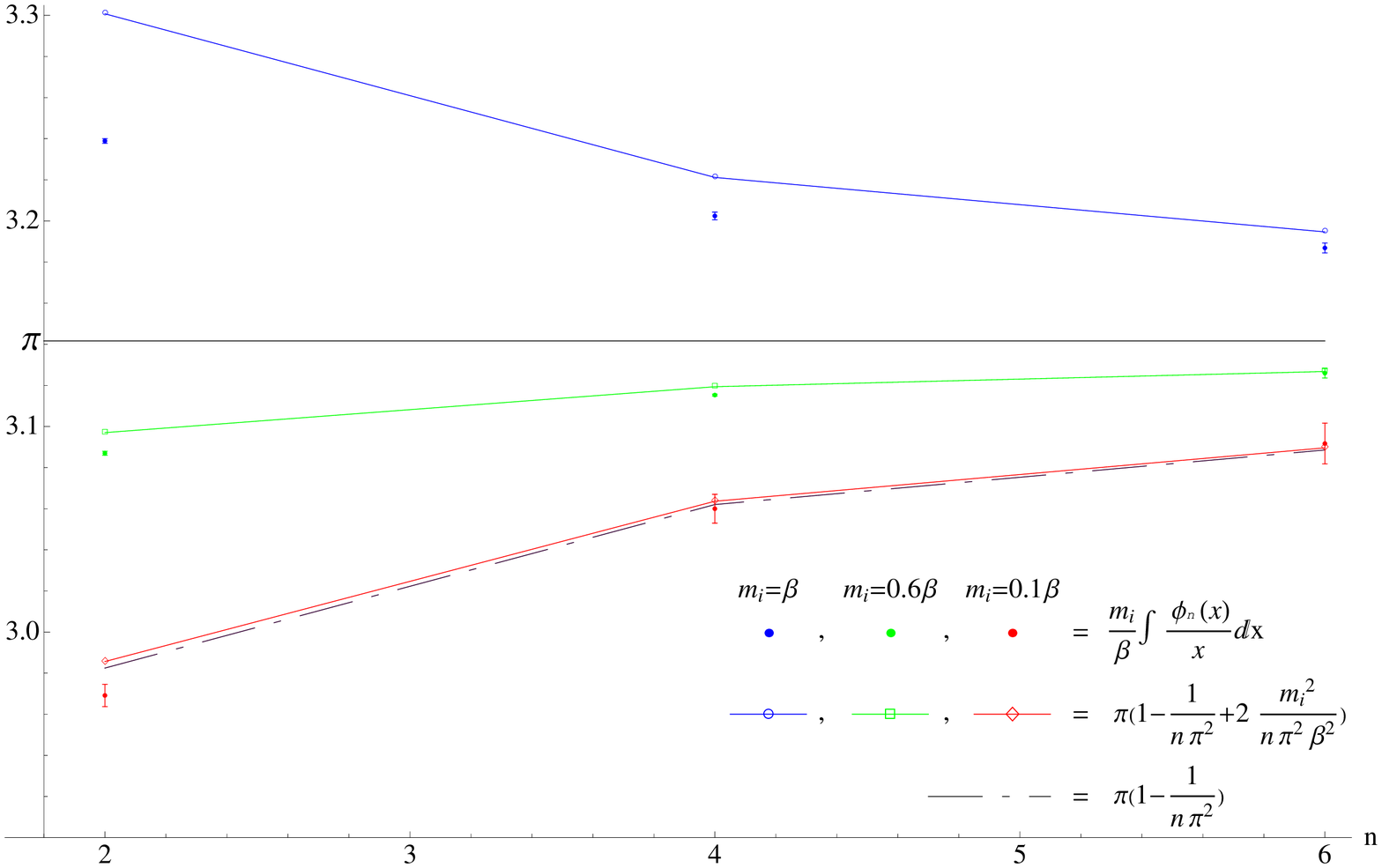}
\caption{\it In this plot we show our results for Eq. (\ref{int1})
with method 2. The bands reflect the difference between using directly Eq. (\ref{int1}) or 
a properly weighted combination of Eq. (\ref{int0}), which we show in tables \ref{table_J1} and 
\ref{table_J2}.}
\label{fig_int1_J}
\end{figure}

We now first try to check Eq. (\ref{int1}). In Fig. \ref{fig_int1_J} we show our results 
with method 2. 
The results from the numerical calculation are presented with some rough estimation 
of their uncertainty, performed by considering the difference between evaluating 
directly Eq. (\ref{int1}) or 
a properly weighted (see tables \ref{table_J1} and 
\ref{table_J2} for the definition) combination of Eq. (\ref{int0}). 
We show in tables \ref{table_J1} and 
\ref{table_J2} the differences for two mass values. 
With this method we cannot go to very large values of $n$, 
nor consider different quark masses or odd values of $n$. Nevertheless, for the 
range of values one can consider with this method the agreement is perfect. 
The $1/n$ and mass dependence can be unambiguously checked very nicely. 

\begin{table}[hhh]
$$
\begin{array}{clclclc}
\hline
n&\vline&\frac{m_i}{\beta}\int dx \frac{\phi_n(x)}{x}&\vline&\frac{M^2(n)}{2m_i\beta}\int dx \phi_n(x)\\
\hline
2&\vline & 3.24 &\vline& 3.24\\
\hline
4 &\vline& 3.20 &\vline& 3.20\\
\hline
6 &\vline& 3.19 &\vline& 3.18\\
\hline
\end{array}
$$
\caption{Results from the numerical calculation using method 2 for $m_i=m_j=\beta$.}
\label{table_J1}
\end{table}
\begin{table}[hhh]
$$
\begin{array}{clclclcc}
\hline
 n&\vline&\frac{m_i}{\beta}\int dx \frac{\phi_n(x)}{x}&\vline&\frac{M^2(n)}{2m_i\beta}\int dx \phi_n(x)\\
\hline
2&\vline & 2.97&\vline & 2.96\\
\hline
4&\vline & 3.06&\vline &3.05\\
\hline
6 &\vline& 3.10 &\vline& 3.08\\
\hline
\end{array}
$$
\caption{Results from the numerical calculation using method 2 for $m_i=m_j=0.1\beta$.}
\label{table_J2}
\end{table}

\begin{figure}[hp]
\centering
\includegraphics[width=1\columnwidth]{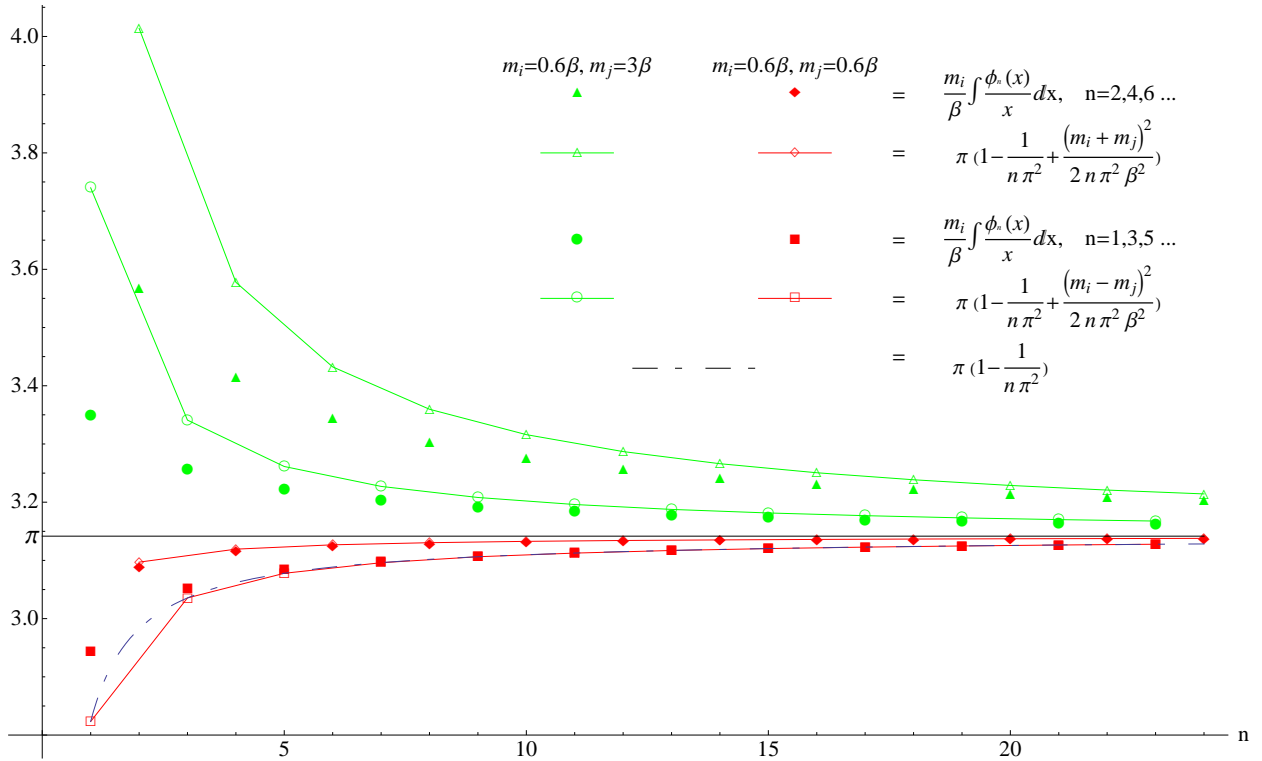}
\caption{\it In this plot we show our results 
with method 1 for Eq. (\ref{int1}). }
\label{fig_int1_sin}
\end{figure}

 In Fig. \ref{fig_int1_sin} we show our results with method 1. 
 In this case we can take much larger 
 values of $n$, although not  so small quark masses as in method 2. On the other 
 hand, in this case, the use of Eq. (\ref{int1}) or 
the properly weighted combination of Eq. (\ref{int0}) is basically indistinguishable. 
We can see clearly in this figure that scalar and pseudoscalar states follow different 
curves, as their corrections involve the difference of the masses of the quarks or their addition, respectively. 
The behaviour of the corrections changes from increasing to decreasing with $n$ depending on how the sum 
and the difference of the quark masses compare to $2\beta$. We can also see how the numerical results 
approach the expected analytic expressions as $n$ increases. 

We conclude that we have been able to unambiguously and nicely check Eqs. (\ref{int1}) and (\ref{int0})
numerically. We note that we have been able to visualize the mass dependence and the sign alternating terms, and that, in particular, we are reaching a numerical precision below 1\%.

\medskip

\begin{figure}[h]
\centering
\includegraphics[width=0.8\columnwidth]{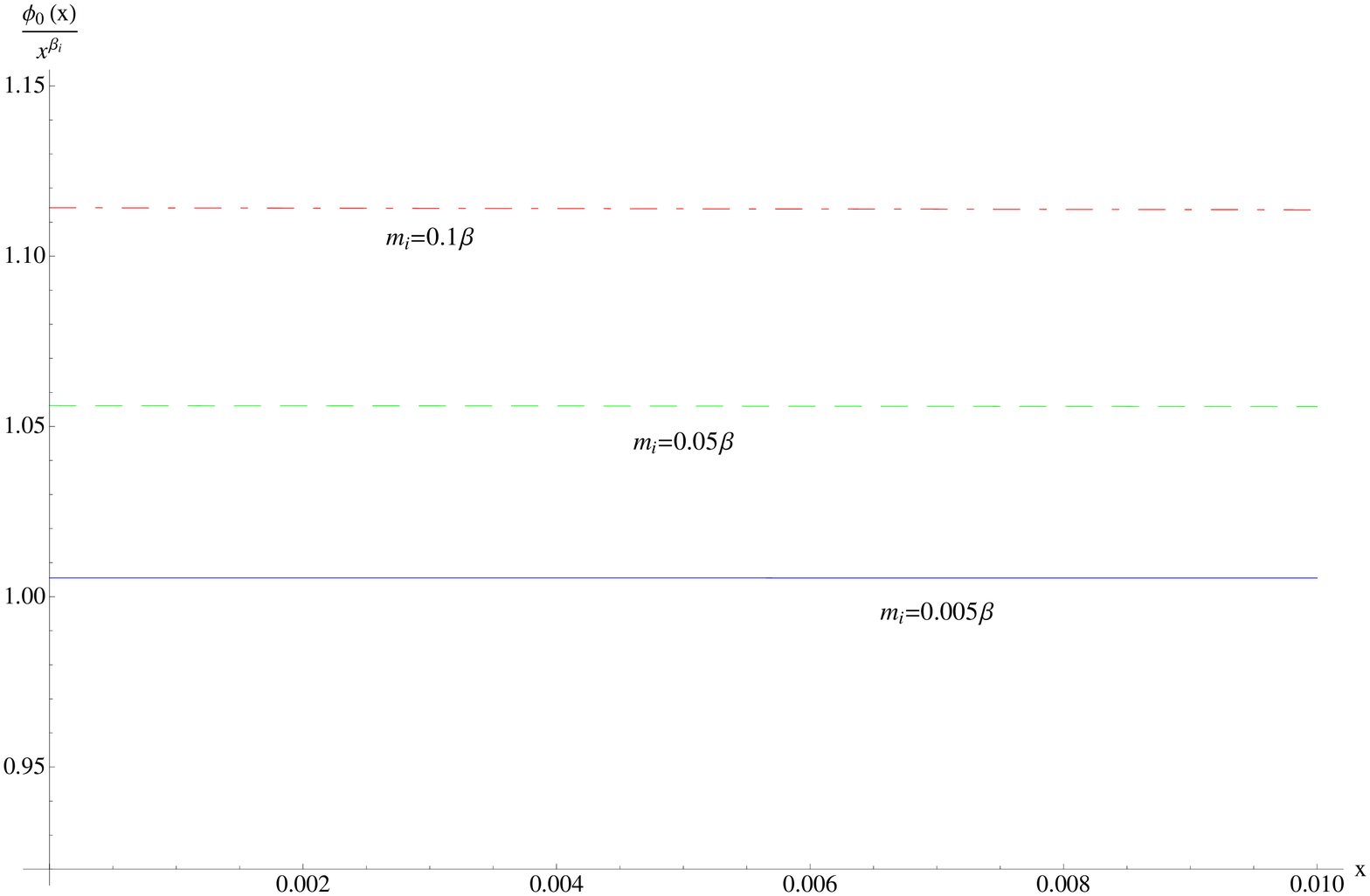}
\caption{\it In this plot we show our results 
with method 2  for Eq. (\ref{c0}) with $m_i=m_j$. 
Due to the small values of the quark masses, 
in the range we are showing the functions are essentially constant.}
\label{fig_c0}
\end{figure}

We now try to check Eq. (\ref{c0}). 
In Fig. \ref{fig_c0} we show the plot of $\phi_0(x)/x^{\beta_i}$ 
for small values of $x$ calculated with method 2, and we can see that, 
indeed, as we approach the massless limit $c_0$ tends to $1$. Method 1 
is not efficient for the evaluation of the wave function near the origin, 
for, as we said before, it does not have the right functional behaviour in 
the $x\to 0,\, 1$ limit: it yields a sinusoidal function that oscillates 
around the right solution, until we approach the boundaries, where the curve that the oscillations follow does not die off as it should.

\medskip

\begin{figure}
\centering
\includegraphics[width=0.8\columnwidth]{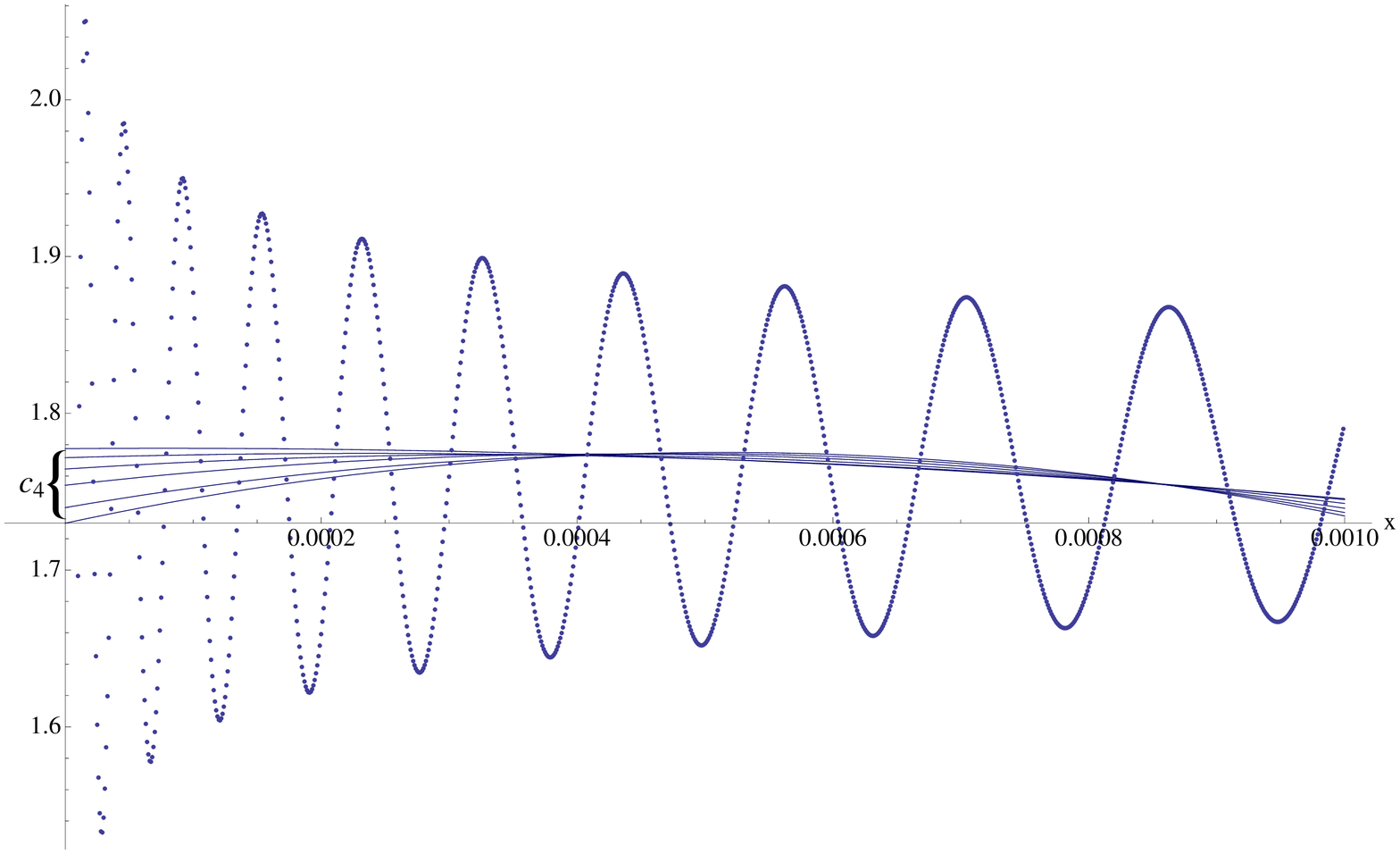}
\caption{\it The dots in this plot correspond to values of $\phi_4(x)/x^{\beta_i}$ sampled with $\Delta x=10^{-6}$ for $m_i=0.001\beta$, $m_j=2\beta$. We show some of the possible fits to a quadratic function, depending on the set of points used, from all those that fall within the interval $x\in [10^{-5},10^{-3}]$ to only those that are within $x\in [10^{-4},10^{-3}]$. The possible values of the fits at the origin give us an estimated range for $c_4$.}
\label{fig_c4}
\end{figure}

\begin{figure}[hh]
\centering
\includegraphics[width=0.8\columnwidth]{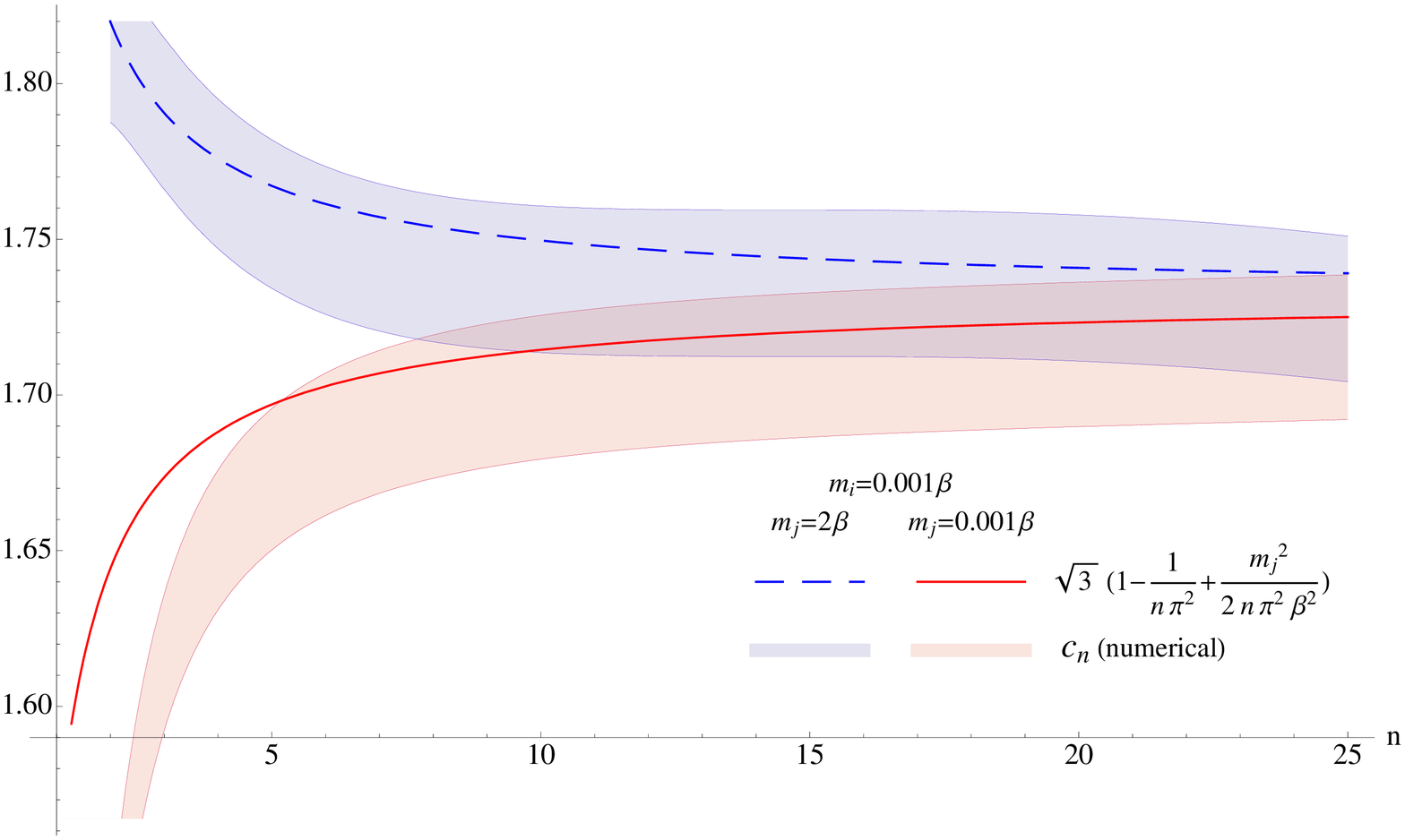}
\caption{\it In this plot we show our results 
with method 1  for Eq. (\ref{cnasympt}). }
\label{fig_cn}
\end{figure}

Finally, we try to check Eq. (\ref{cnasympt}). In this case we cannot use method 2: 
although it incorporates by definition the right $c_nx^{\beta_i}$ behaviour, its 
numerical precision at high values of $n$ is very poor. We are left then with method 1. 
Numerical precision in this case is not so much of an issue, but due to the aforementioned 
limitations of this method we can only hope to get some estimate for $c_n$ by finding the 
path around which our sinusoidal solutions oscillate when $x$ is not too close to $0$, and 
then extrapolate this path to the limit $x\to 0$. In figure \ref{fig_c4} we show our results 
for n=4 with $m_i=0.001\beta$, $m_j=2\beta$: we evaluate $\phi_4(x)/x^{\beta_i}$ at points 
separated a distance $\Delta x=10^{-6}$, starting at $x=10^{-5}$ and ending at $x=10^{-3}$, 
and we fit them to a quadratic function. Depending on how many of the points that are closer 
to the origin we include in the fit, the resulting curve will be one or another, as shown in 
the plot, somewhat oscillating around the two extreme curves displayed. 
The possible values of the fit at $x=0$ are the possible values of $c_4$ that we can 
find with method 1. In figure \ref{fig_cn} we show the results up to $n=25$ for the cases 
$m_i=0.001\beta$, $m_j=2\beta$ and $m_i=0.001\beta$, $m_j=0.001\beta$. Overall, this numerical 
analysis serves as a consistency check of Eq. (\ref{cnasympt}).

\section{Finite $N_c$}

This far the discussion has been restricted to the large $N_c$ limit. 
We would like to finish this paper by studying models valid at 
finite $N_c$. One model with the right analytic properties for the 
current-current correlator and with the correct limit in the large $N_c$ 
limit 
has been built in Ref. \cite{Blok:1997hs}. This model is also able 
to reproduce the leading parton-model logarithm at large $Q^2$. 
Working analogously to previous sections, we will consider whether 
subleading perturbative corrections  
can be incorporated in this model, and see if by including $1/N_c$ effects there appears any connection 
between the slopes of resonances of different parity. 
For definiteness, we will consider the vector-vector correlator 
but most of the discussion trivially applies to any other current-current correlator; 
in particular the scalar or pseudoscalar correlator, 
for which the only complication would be to take into account the anomalous dimensions. 
Therefore, we take
\be
\Pi_V^{\mu\nu}(q)
\equiv
(q^{\mu}q^{\nu}-g^{\mu\nu}q^2)\Pi_V(q)
\equiv
i\int d^4x e^{iqx}\langle vac|J_V^{\mu}(x)J_V^{\nu}(0)|vac\rangle
\,,
\ee
where $J_V^{\mu}=\sum_fQ_f\bar \psi_f\gamma^{\mu}\psi_f$.
In order to avoid divergences, we will work with the Adler function 
\be
{\cal A}_V(Q^2)\equiv -Q^2{d \over dQ^2}\Pi_V(Q^2)=
Q^2\int_0^{\infty}dt \frac{1}{(t+Q^2)^2}\frac{1}{\pi}\text{Im}\Pi_V(t)
\,.
\ee

We consider the following expression for the 
vacuum polarization,
\be
\Pi_V(Q^2)
=
\sum_{n=0}^{\infty}
{F_V^2(n) \over \left(\frac{Q^2}{\Lambda^2}\right)^{1-\frac{C}{\pi N_c}}\Lambda^2+M_V^2(n)}
\,.
\ee
For $F_V^2(n)=$constant and $M_V^2(n)=B_Vn$, we recover the model of Ref. \cite{Blok:1997hs}. 
Here we will allow $F_V^2(n)$ to be $n$-dependent. 
The Adler function for this model reads
\be
{\cal A}_V(Q^2)=(1-\frac{C}{\pi N_c})z\sum_{n=0}^{\infty}\frac{F_V^2(n)}{\Lambda^2}\frac{1}{(z+n)^2}
\,, \qquad
z=\left(\frac{Q^2}{\Lambda^2}\right)^{1-\frac{C}{\pi N_c}} \ .
\ee
For simplicity we will set 
$\Lambda^2=B_V$ in the following. It will also be enough for our purposes to keep the analysis of ${\cal A}(Q^2)$ to the 
lowest order in the OPE. Therefore, we only consider
\bea
\label{pertV}
{\cal A}_{V,pert.}(Q^2)
=
{\cal A}_0\left(
1+c_1\frac{\alpha_{\cal A}(Q^2)}{\pi}
\right)
=
-Q^2{d \over dQ^2}
{\cal A}_0
\int_0^{\infty}dt\frac{1}{(t+Q^2)^2}
\left(
1+c_1\frac{\alpha_{M}(t)}{\pi}
\right)
\label{Apert}
\,,
\eea
where $\alpha_{\cal A}(Q^2)$ and $\alpha_{M}(t^2)$ 
admit an analytic expansion in terms 
of $\als(Q^2)$ (computed in the $\MS$ scheme): 
\bea
\alpha_{\cal A}(Q^2)&=&\als(Q^2)\left(1+\frac{c_2}{c_1}\frac{\als(Q^2)}{\pi}
+\frac{d_3}{c_1}
\frac{\als^2(Q^2)}{\pi^2}+\cdots
\right) \\
\alpha_{M}(t)&=&
\als(t)\left(1+\frac{c_2}{c_1}\frac{\als(t)}{\pi}+\frac{c_3}{c_1}
\frac{\als^2(t)}{\pi^2}+\cdots 
\right) \ .
\eea
The coefficients $c_i$ and $d_i$ have been computed in Ref. \cite{Chetyrkin:1996}
(they can be obtained from each other through dispersion relations):
\bea
c_1&=&\frac{3N_c}{8}\\
c_2&=&\frac{N_c^2}{128}\left((243-176\zeta(3))
-\frac{4 n_f}{N_c}(11-8\zeta(3))\right)\nn
\\ 
\nn
c_3&=&\frac{N_c^3}{27648}\left[346201-2904\pi^2-324528\zeta(3)+63360\zeta(5) \nn \right.\\
&&\left.+\frac{2 n_f}{N_c}(-62863+528\pi^2+51216\zeta(3)-5760\zeta(5))\right]
+\frac{\left(\sum_fQ_f\right)^2}{\sum_fQ_f^2}\frac{N_c^2}{1024}\left(\frac{176}{3}-128\zeta(3)\right) \ ,\nn 
\eea
and 
$$
{\cal A}_0
=
\sum_fQ_f^2
\frac{4}{3}\frac{N_c}{16\pi^2} \ .
$$
As before, through the Euler-Maclaurin formula we can see that the matching between the perturbative and 
the hadronic calculations at logarithmic order requires that
\be
{\cal A}_0\left(
1+c_1\frac{\alpha_{\cal A}(Q^2)}{\pi}
\right)
=
\left(1-\frac{C}{\pi N_c}\right) z \int_0^{\infty}dn \frac{F_{V,0}^2(n)}{B_V}\frac{1}{(z+n)^2} \ ,
\ee
where we have added the subscript $0$ to $F_V^2(n)$, in agreement with our notation of sec. \ref{matching}.
In Ref. \cite{Mondejar:2007dz}, we obtained that 
\be
\frac{F_{V,0}^{2,\infty}}{B_V}
=
{\cal A}_0
\left(
1+c_1^{\infty}\frac{\alpha^{\infty}_{M}(B_Vn)}{\pi}
\right) \ ,
\ee
where the upperscript $\infty$ stands for the $N_c \rightarrow \infty$ limit. 
 One could consider that the same structure should hold 
for finite $N_c$ in our model and that the decay constant should then be
proportional to $1+c_1\frac{\alpha_{M}(B_Vn)}{\pi}$. 
However, this is not so due to the specific functionality on 
$Q^2$ that we have introduced. For instance, it is 
interesting to consider the following quantity
\be
z\int_0^{\infty}dn\frac{1}{(n+z)^2}\alpha_{M}(B_Vn)=\alpha_{\cal A}(B_Vz)
=
\alpha_{\cal A}(Q^2)-\frac{C}{\pi N_c}\ln\frac{Q^2}{B_V}\beta_{\cal A}^{\infty}(\alpha_{\cal A}^{\infty}(Q^2))
+O(\frac{1}{N_c^2}) \ .
\ee
The first term is what we should obtain to match the perturbative calculation. If we want $F_{V,0}^2(n)$ to be expressed in terms of $\als(B_Vn)$ we 
have to fine tune it in order to eliminate the second extra 
term.

If we write 
\be
\frac{F_{V,0}^{2}}{B_V}
=
\frac{{\cal A}_0}{1-
\frac{C}{\pi N_c}}\left(
1+c_1\frac{\alpha_{M}(B_Vn)}{\pi}
\right)
+
\frac{C}{\pi N_c}\frac{\delta F_{V,0}^{2}}{B_V} \ ,
\ee
the following equality has to be satisfied (here, with the precision of the 
calculation, we can replace $z\simeq Q^2/B_V$)
\be
z\int_0^{\infty}dn\frac{1}{(n+z)^2}\frac{\delta F_{V,0}^{2}}{B_V}
=
\ln\frac{Q^2}{B_V}\beta_{\cal A}^{\infty}(\alpha_{\cal A}^{\infty}(Q^2)) \ .
\ee
This implies that $\delta F^2 \sim \als^2(B_V)\ln n$. 
Note that $\beta_0$ and $\beta_1$ are the same for $\alpha_{\cal A}$, 
$\alpha_{M}$ and $\als$. Therefore, at low orders in $\alpha$ the distinction between 
different renormalization schemes is superfluous. 

We can try to refine our ansatz. If  we rewrite the 
expression for $F_V^2$ in the following way,
\bea
\frac{F_{V,0}^{2}}{B_V}
&=&
\frac{{\cal A}_0}{1-
\frac{C}{\pi N_c}}\left[\left(
1+c_1\frac{\alpha_{M}(B_Vn^{1+\frac{C}{\pi N_c}})}{\pi}
\right)
+
\frac{C}{\pi N_c}\frac{\delta {\tilde F}_{V,0}^{2}}{B_V}\right]
\\
\nn
&=&
\frac{{\cal A}_0}{1-
\frac{C}{\pi N_c}}\left[\left(
1+c_1\frac{\alpha_{M}(B_Vn)}{\pi}+
\frac{c_1}{\pi}\frac{C}{\pi N_c}\ln n \beta_{M}^{\infty}(\alpha_{M}^{\infty}(B_Vn))
+O(\frac{1}{N_c^2})
\right)
+
\frac{C}{\pi N_c}\frac{\delta {\tilde F}_{V,0}^{2}}{B_V}\right] \ ,
\eea
some logs are reabsorbed in $\alpha_{M}(B_Vn^{1+\frac{C}{\pi N_c}})$ and  
$\delta {\tilde F}_{V,0}^2 \sim {\cal O}(\als^3(B_V), \als^{3+s}(B_V)\ln^s n)$.
We can easily obtain the coefficient multiplying $\als^3(B_V)$\footnote{Nevertheless, 
in this case it is more difficult to get a closed expression with $\als^{3+s}(B_V)\ln^s n$ 
accuracy.
We do not undertake the effort to try to get it. At this level of precision our expression 
is good enough and also good enough to make our point.}.
We find that, at $\mathcal{O}(\als^3)$,
\be
\frac{\delta {\tilde F}_{V,0}^2}{B_V}= -\frac{2}{3\pi}\beta_0^2 c_1 \als(B_V)^3 \ .
\ee

The calculations for the axial-vector case are equal replacing $B_V$ by $B_A$ (the slope of the 
Regge trajectory in the axial-vector case). 
Therefore, as in Ref. \cite{Mondejar:2007dz}, we can satisfy the constraints of the OPE with  
independent values for $B_V$ and $B_A$. Thus, we conclude that the OPE cannot constrain
$B_V$ and $B_A$ to be equal for general models with the right analytic properties at finite 
$N_c$. Note as well that, in general, at finite $N_c$
\be
\frac{F_{V,0}^2(n)}{|dM_{V,LO}^2(n)/dn|} \not= \frac{1}{\pi}{\rm Im}\Pi^{pt.}_V(M_{V,LO}^2(n))
\,.
\ee
 
\subsection{$\rho$ at finite $N_c$}
 
We now briefly consider what this model tells us about 
the computation of $\rho(t)\equiv \frac{1}{\pi}{\rm Im} \Pi_V(t)$. This is 
a physical observable related with $\sigma(e^+e^- \rightarrow {\rm hadrons})$. 

We first note that one can approximate the sum over $n$ by an integral 
(i.e. use the Euler McLaurin expansion) if $q^2 \rightarrow \infty$ and $|\arg z| < \pi$  for 
the model of Ref. \cite{Blok:1997hs}.  At this respect the modification to this model introduced by 
us is simply the introduction of logs of $n$ 
in the decay constants, which should not change the analytic properties of the function. 
This allows us to obtain the OPE expression for the vacuum polarization from the hadronic expression 
if $|\arg z| < \pi$, as in the original model of Ref. \cite{Blok:1997hs}.
For $q^2$ positive we do have $|\arg z| < \pi$. Thus, we can find the OPE expression from the hadronic one, 
and vice versa, obtain the hadronic expression from the OPE. In particular, we can obtain its imaginary part, 
which corresponds to $\rho(t)$. 
Part of the original structure of $\rho(s)$ will be lost in the way, as the OPE misses non-analytic terms. The associated error
is exponential with an exponent that vanishes when $n_f \rightarrow 0$ (leaving aside the fact that one expects 
the series to be asymptotic).
Therefore, the general structure of the solution would be 
$$
\rho(s)=\rho_{pert.}(s)+ ({\rm power\ like}) + ({\rm exp.\; suppr.})
\,,
$$
where only the last term does not follow from the OPE. Unfortunately this 
result has been obtained for an specific model. In general, one cannot get 
in a model independent way that 
$$
\rho(s)=\rho_{pert.}(s)+ \cdots
\,,
$$
where the dots mean a contribution that 
decays to zero faster than any power of $1/\ln s$. 

\section{Conclusions}

We have studied the constraints that the OPE imposes on models for current-current 
correlators inspired in the large $N_c$ limit and Regge trajectories. 
We have considered the case of scalar/pseudoscalar correlators. 
Assuming a model for the mass spectrum consisting of a linear Regge behaviour plus corrections 
in $1/n$, we have obtained the logarithmic behaviour in $n$ of the decay constants 
within a systematic expansion in $1/n$. We have accomplished this by matching 
the hadronic and OPE expressions of the Adler function. The inclusion of $1/n$ 
corrections to the decay constants is compulsory if one aims at going beyond the 
leading partonic result, as the $1/n$ terms are needed to produce logarithms of $Q^2$ 
in the Euclidean. We have performed a numerical analysis of our results, in which we 
have seen the importance of the resummation of the $\ln n$ terms of 
the decay constants, specially for low $n$. Our results show that it is possible to have a different slope, $B_{S/P}$, for 
the Regge behaviour of the scalar and pseudoscalar spectrum, and yet comply with all the constraints 
imposed  by the OPE (including perturbative corrections).

In order to check our setup in a controlled environment, where the results obtained with the 
OPE can be tested with the exact results, we have studied the preasymptotic effects of the 
scalar/pseudoscalar correlators in the 't Hooft model (QCD in 2 dimensions in the large $N_c$ limit). 
This has allowed us to compute the $1/n$ corrections to the decay constants in the 't Hooft model 
for the first time. Actually, the connection between the OPE and the decay constants provides us 
with a relatively easy way of finding $1/n$ corrections to hadronic matrix elements in the 't Hooft model.
A direct analytic computation from the 't Hooft equation 
appears to be quite involved, however we have been able to confirm our results with the numerical evaluation from the 
't Hooft equation. 

Finally, we have also considered a model at finite $N_c$ 
\cite{Blok:1997hs} and modified it to allow for the inclusion of 
perturbative (logarithmic) corrections. 
We have then obtained the associated decay constants that are consistent with 
perturbation theory, though they are no longer trivially related with the 
perturbative expression of the imaginary part of the correlator.
We have seen that consistency of this model with the OPE can be obtained with 
different slopes for the Regge trajectory associated to each channel.  

\medskip

{\bf Acknowledgments}. This work is partially supported by the 
network Flavianet MRTN-CT-2006-035482, by the spanish 
grant FPA2007-60275, by the Spanish Consolider-Ingenio 2010
Programme CPAN (CSD2007-00042), and by the catalan grant SGR2005-00916.

\appendix 

\section{Appendix}
In this appendix we will present general, model-independent, matching formulas 
between the OPE and the hadronic calculation both for vector/axial-vector (for the 
notation in this case, see Ref. \cite{Mondejar:2007dz}) and 
for scalar/pseudo-scalar correlators, within an expansion in powers of $1/M_n^2$.

The procedure is the same as in the main body of the text. 
Take the Adler function ($\mathcal{A}_X$ in the vector/axial-vector case, 
$\mathcal{B}_X$ in the scalar/pseudo-scalar case), transform it using the 
Euler-Maclaurin formula, and focus on the piece that can produce logarithms, 
namely, the integral. 

\subsection{Vector/Axial-vector correlators}

In this case we work with
\be
Q^2 \int_{n^*}^{\infty} dn \frac{F^2_X(n)}{(Q^2 +M_n^2)^2} = Q^2 \int_{M_{n^*}^2}^{\infty} dM_n^2 \frac{1}{\left|\frac{dM_n^2}{dn}\right|}\frac{F^2_X(n)}{(Q^2 +M_n^2)^2} \ .
\ee
Now, instead of assuming a model for $M_n$, we will just expand
\be
\frac{1}{\left|\frac{dM_n^2}{dn}\right|}F^2_X(n) \equiv \widetilde{F}_{X,0}(M_n^2)
+ \frac{\widetilde{F}_{X,1}(M_n^2)}{M_n^2}+ \frac{\widetilde{F}_{X,2}(M_n^2)}{M_n^4} + \dots
\ee
Within this expansion, retracing the steps of section 3 in Ref. \cite{Mondejar:2007dz} the matching conditions between OPE 
and hadronic calculations are trivially obtained:
\begin{itemize}
\item{LO: $\widetilde{F}_{0,X}(M_n^2)= \text{\Large{$\frac{1}{\pi}$}} \text{Im} \Pi^{pt.}_X(M_n^2)$}
\item{NLO: $\widetilde{F}_{1,X}(M_n^2)=0$}
\item{NNLO: $\widetilde{F}_{2,X}(M_n^2)=\text{\Large{$\frac{35}{968\pi^2}$}}\beta_0 N_c^2\als^2(M_n^2)
\text{\Large{$\frac{\beta(\als(\mu))\langle vac|G^2(\mu)|vac\rangle}{N_c^2}$}}\ .$}
\end{itemize}

\subsection{Scalar/Pseudo-scalar correlators}

Here we work with
\be
Q^4 \int_{n^*}^{\infty} dn \frac{F^2_X(n)}{(Q^2 +M_n^2)^3} = 
Q^4 \int_{M_{n^*}^2}^{\infty} dM_n^2 \frac{1}{\left|\frac{dM_n^2}{dn}\right|}\frac{F^2_X(n;\mu)}{(Q^2 +M_n^2)^3} \ .
\ee
As the dimensions of these correlators are different from those of the vector/axial-vector correlators, we will use a different expansion,
\be
\frac{1}{\left|\frac{dM_n^2}{dn}\right|}F^2_X(n;\mu) \equiv M_n^2\lp \bar{F}_{X,0}(M_n^2;\mu)+ 
\frac{\bar{F}_{X,1}(M_n^2;\mu)}{M_n^2}+ \frac{\bar{F}_{X,2}(M_n^2;\mu)}{M_n^4} + \dots\rp \ .
\ee
And the matching results are
\begin{itemize}
\item{LO: $\bar{F}_{0,X}(M_n^2;\mu)=\text{\Large{$\frac{1}{M_n^2}\frac{1}{\pi}$}} \text{Im} \Pi^{pt.}_X(M_n^2;\mu)$}
\item{NLO: $\bar{F}_{1,X}(M_n^2;\mu)=0$}
\item{NNLO: $\bar{F}_{2,X}(M_n^2;\mu)= \text{\Large{$\frac{3}{11\pi}$}}\beta_0\bar{\gamma_0}
\text{\large{$\lp \frac{\als(M_n^2)}{\als(\mu^2)}\rp^{2\bar{\gamma_0}}$}}N_c\als(M_n^2)\text{\Large{$\frac{\beta(\als(\mu))
\langle vac|G^2(\mu)|vac\rangle}{N_c^2}$}}$} \ .
\end{itemize}


\end{document}